\documentclass[12pt,preprint]{aastex}
\newcommand{\bzeta}{\mbox{\boldmath{$\zeta$}}}
\newcommand{\bgamma}{\mbox{\boldmath{$\gamma$}}}
\newcommand{\balpha}{\mbox{\boldmath{$\alpha$}}}
\newcommand{\bu}{\mbox{\boldmath{$u$}}}
\newcommand{\bq}{\mbox{\boldmath{$q$}}}
\newcommand{\bov}{\mbox{\boldmath{$v$}}}

\newcommand{\bxi}{\mbox{\boldmath{$\xi$}}}
\newcommand{\boeta}{\mbox{\boldmath{$\eta$}}}
\newcommand{\bx}{\mbox{\boldmath{$x$}}}
\newcommand{\br}{\mbox{\boldmath{$r$}}}
\newcommand{\bl}{\mbox{\boldmath{$l$}}}
\newcommand{\tr}{\textrm{tr}}

\shorttitle{Lensing Effects on Gravitational Waves in a Clumpy Universe}
\shortauthors{Yoo et al.}

\begin{document}

\title{Lensing Effects on Gravitational Waves in a Clumpy Universe\\
---{\em  Effects of Inhomogeneity on the Distance-Redshift Relation}---}

\author{Chul-Moon Yoo\footnote{E-mail:c\_m\_yoo@sci.osaka-cu.ac.jp}
, 
Ken-ichi Nakao\footnote{E-mail:knakao@sci.osaka-cu.ac.jp} 
and
Hiroshi Kozaki\footnote{E-mail:furusaki@sci.osaka-cu.ac.jp}
}

\affil{%
Department of Mathematics and Physics, Graduate School of Science,
Osaka City University, Osaka 558-8585, Japan
}

\author{Ryuichi Takahashi\footnote{E-mail:Takahasi@th.nao.ac.jp}
}

\affil{%
Division of Theoretical Astronomy,
National Astronomical Observatory of Japan, 
Mitaka, Tokyo 181-8588, Japan
}

\begin{abstract}
The distance-redshift relation determined by means of 
gravitational waves in the 
clumpy universe is simulated numerically by taking into account 
the effects of gravitational lensing.
It is assumed that all of the matter in the universe takes the form of 
randomly distributed point masses, each of which has the identical mass
$M_L$.
Calculations are carried out in two extreme cases:
$\lambda\gg GM_L/c^2$ and $\lambda\ll GM_L/c^2$,
where $\lambda$ denotes the wavelength of gravitational waves.
In the first case, the distance-redshift relation for the fully
homogeneous and isotropic universe is reproduced with a small 
distance dispersion, whereas in the second case, the distance 
dispersion is larger. This result suggests that we might obtain 
information about the typical mass of lens objects through the  
distance-redshift relation gleaned through observation of gravitational 
waves of various wavelengths. 
In this paper, we show how to set limitations on the 
mass $M_L$ through the observation of gravitational waves 
in the clumpy universe model described above.

\end{abstract}



\clearpage

\slugcomment{
\hspace{-5cm}\baselineskip0pt
\leftline{\large\baselineskip16pt\sl\vbox to0pt{\hbox{\em Department of
Mathematics and Physics}
               \hbox{\em Osaka City  University}\vss}}

\rightline{\large\baselineskip16pt\rm\vbox to20pt{\hbox{OCU-PHYS-244}
            \hbox{AP-GR-32}
\vss}}%

\vspace{1cm}}

\section{Introduction}\label{sec:intro}
Owing to recent developments of observational techniques, 
it is expected that detection of gravitational waves 
will be achieved in the near future.
Laser interferometer gravitational wave detectors, 
such as the Tokyo Advanced Medium-Scale Antenna~(TAMA-300), 
the Laser Interferometer Gravitational-Wave Observatory~(LIGO), 
Variability of Irradiance and Gravity Oscillations~(VIRGO)
and GEO-600, are in operation, and other projects, 
such as the Large-scale Cryogenic Gravitational wave Telescope~(LCGT), 
LIGO2, the {\em Laser Interferometer Space Antenna~(LISA)},
{\em the Decihertz Interferometer Gravitational Wave Observatory}~({\em DECIGO};
\cite{Seto:2001qf}), and the {\em Big Bang Observer~(BBO)},
are in the planning stage. 
In preparation for future breakthroughs associated with these projects, 
it is useful to discuss the information provided by gravitational wave data.
Here we focus on inhomogeneities of our universe that can be
investigated with gravitational wave observations. 

Many cosmological observations suggest that our universe is globally
homogeneous and isotropic, in other words,
that the time evolution of the global aspect is well approximated by 
the Friedmann-Lema\^itre(FL) cosmological model.
However, our universe is locally inhomogeneous.
The effects of these inhomogeneities on the evolution of our universe or 
on cosmological observations draw our attention as one of the most
important issues in cosmology.
However, it is difficult to directly observe the inhomogeneities with
optical observations because most of the matter in our universe is not
luminous.
One useful means of obtaining information about aspects of 
inhomogeneities is to examine gravitational lensing. 
Gravitational and electromagnetic waves are subject to gravitational
lensing effects caused by inhomogeneously distributed matter around
their trajectories. 
Therefore, we may find features of inhomogeneities through  
both observational and theoretical studies of the 
gravitational lensing effects in  
astronomically significant situations.  

There are various candidates for the dark matter such as weakly interacting
massive particles (WIMPs) and massive compact halo objects(MACHOs). 
Several observations have been undertaken to constrain the 
mass density of compact objects, $\Omega_{\rm CO}$; 
direct searches for MACHOs in the Milky Way have been 
performed by the MACHO and EROS collaborations 
through microlensing surveys. 
The MACHO group~\citep{Alcock:2000ph} group concluded that 
the most likely halo fraction in form of compact 
objects with masses in the range 0.1-1$M_\odot$ is 
of about 20\%. The EROS team concluded that the objects in the mass range 
from $2\times 10^{-7}M_\odot$ to $1M_\odot$ cannot 
contribute more than 25\% of the total halo~\citep{Afonso:2003}. 
However, the universal fraction of macroscopic dark matter 
could be significantly different from these local estimates. 
Millilensing tests of gamma-ray bursts 
derive a limit on $\Omega_{\rm CO}$ of 
$\Omega_{\rm CO}<0.1$ in the mass range from $10^5$ to 
$10^9M_\odot$~\citep{Nemiroff:2001}. 
Multiple imaging searches in compact radio sources 
derive the limit on $\Omega_{\rm CO}$ as 
$\Omega_{\rm CO}\lesssim0.013$ in the mass range 
from $10^6$ to $10^8M_\odot$~\citep{Wilkinson:2001}. 
The mass range of these observational tests are 
limited, and thus other methods to investigate the mass fraction of
macroscopic compact objects in wider mass range are needed. 

In this paper, we show that it is possible to 
extract information about the properties of macroscopic 
compact objects from the observational data of 
gravitational waves by analyzing the gravitational lensing effects due to 
these compact objects. 
For this purpose, we consider an idealized 
model of the inhomogeneous universe which has the following properties.
First, this model is a globally FL universe. 
Second, all of the matter takes the form of point masses, 
each of which has the identical mass $M_L$.
Finally, the point masses are uniformly distributed. 

One of the main targets of gravitational-wave astronomy is 
binary compact objects. 
Their gravitational waves have much longer wavelengths $(\lambda)$ than 
the optical, and furthermore, these are coherent. 
In typical optical observations, the wavelength 
$(\lambda \sim1 \mu{\rm m})$ might be much shorter than the
Schwarzschild radii of lens objects.\footnote{It does not conflict 
with any observational result that there might be 
lens objects of Schwarzschild radii much smaller than $1\mu{\rm m}$. 
However, for example, supernovae which might be the smallest 
and yet very bright optical sources, are larger than the Einstein radii 
of such small-mass lens objects, and thus the lensing effects due to 
those will be negligible. }
Thus, we usually analyze the gravitational lensing effects on 
electromagnetic waves by using geometrical optics. 
In contrast, we need wave optics for the gravitational lensing 
of gravitational waves, since the wavelength of gravitational waves may be 
comparable to or longer than the Schwarzschild radii of lens objects. 
For example, the wave-band of {\em LISA} is typically
$10^{11}-10^{14}$ cm; a point mass of
$10^5 - 10^8 M_{\odot}$ has a Schwarzschild radius almost equal to the 
wavelengths within that wave-band. 
When we consider gravitational lensing effects on the gravitational 
waves with $GM_L / c^2 \lesssim \lambda$, we have to take the wave effect 
~\citep{Nakamura:1997sw} into account. 
In fact, remarkable differences between the extreme cases 
$\lambda \gg GM_L / c^2$ and $\lambda \ll GM_L/c^2$ 
have already been reported~\citep{Takahashi:2003ix}.

In order to obtain information about uniformly distributed 
compact objects by investigating 
the gravitational lensing effects in observational data, 
we focus on the relation between the distance from the observer to the source 
of the gravitational waves and its redshift. 
The distance is determined using the 
information contained in the amplitudes of the gravitational waves 
from the so-called standard sirens~\citep{Holz:2005df}. 
Although there are several similar analyses using the 
optical observation of Type Ia supernovae~\citep{Holz:2004xx}, 
gravitational lensing effects on 
gravitational waves can add new information about 
the properties of the lensing compact objects 
to that obtained by optical observations, by virtue of the wave effect, 
through which 
we can gain an understanding of the typical mass and number density. 

Here we assume that the redshift of each source is independently given
by the observation of the electromagnetic
counterpart~\citep{Kocsis:2005vv} or the waveform of the gravitational
waves~\citep{Markovic:1993cr}.
The luminosity distance $d_{\rm l}$ from the observer 
to the source is given by the
waveform of the gravitational waves, as follows~\citep{schutz86}.
Since the orbit of binary will be quickly circularized by a gravitational 
radiation reaction~\citep{Peters:1963,Peters:1964}, the effect of 
ellipticity is negligible when the emitted gravitational 
radiation becomes so strong that we can observe it. Then, the 
two wave modes emitted from a binary of two compact objects with 
masses $m_1$ and $m_2$ are given by 
\begin{eqnarray}
 h_+
 &=& \frac{2 \mathcal{M}_{\rm chirp}^{5/3} (\pi f)^{2/3} \cos(2 \pi ft)}
          {d_{\rm l}} (\cos^2 \theta + 1),
 \label{eq:plus}\\
 h_{\times}
 &=& \frac{2 \mathcal{M}_{\rm chirp}^{5/3} (\pi f)^{2/3} \sin(2 \pi ft)}
          {d_{\rm l}} 2 \cos \theta, 
 \label{eq:cross}
\end{eqnarray}
where $f$ and $\theta$ are the frequency and the angle between the
angular momentum vector of the binary and the line-of-sight, respectively, 
and $\mathcal{M}_{\rm chirp}$ is the redshifted chirp mass, defined by 
\begin{equation}
 \mathcal{M}_{\rm chirp}
 = (1 + z) M_{\rm chirp}
 = (1 + z) \frac{(m_1 m_2)^{3/5}}{(m_1 + m_2)^{1/5}}. 
\end{equation}
Assuming a circular orbit, we have 
\begin{equation}
 \frac{df}{dt}
 = \frac{96 \pi^{8/3}}{5} f^{11/3} \mathcal{M}_{\rm chirp}^{5/3}.
 \label{eq:fdot}
\end{equation}
Eliminating $\mathcal{M}_{\rm chirp}$ and $\theta$ from
equations~(\ref{eq:plus}), (\ref{eq:cross}), and (\ref{eq:fdot}),
we can obtain $d_{\rm l}$ if there is no gravitational lensing effect 
on the gravitational waves. However, if the gravitational waves 
are gravitationally lensed, their wave forms are changed.
The gravitational lensing effects may lead to an
incorrect estimation of $d_l$, and thus 
we have to know how the gravitational lensing changes the waveforms of 
the gravitational waves.

This paper is organized as follows. 
In \S\ref{sec:geo}, the basic theory of gravitational lensing is
introduced. We discuss the lensing probability in \S\ref{sec:dispro}. 
Then we show the calculation method and results for the distance-redshift
relation in \S\S\ref{sec:wlim} and \ref{sec:glim}.
Finally, \S\ref{sec:consum} is devoted to the conclusions and summary.
Throughout the paper, we adopt the unit of $G=c=1$.

\section{Review of gravitational lensing}\label{sec:geo}
In this section, we introduce basic equations for gravitational lensing.

\subsection{Geometrical optics}
\begin{figure}[htbp]
 \begin{center}
  \includegraphics[scale=0.8]{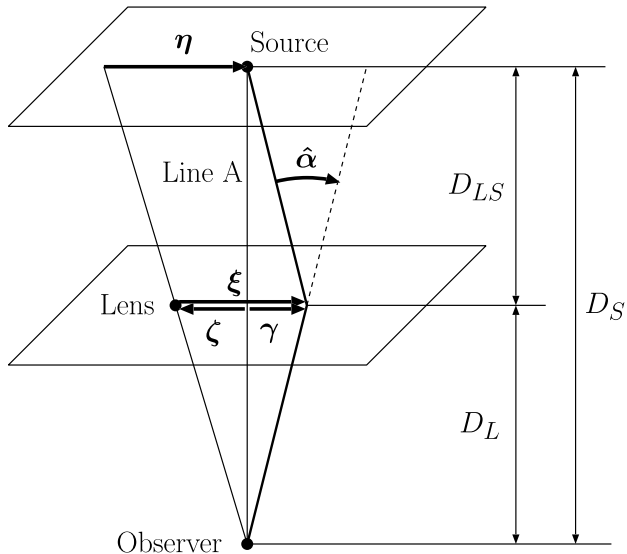}
  \caption{Geometry of gravitational lensing by a point-mass lens.
  The impact vector $\bxi$ represents the relative position of 
  a light ray on the lens plane to the lens position, and 
  $\hat{\balpha}$ is a vector whose norm is equal to 
  the bending angle.} 
  \label{fig:lens}
 \end{center}
\end{figure}

In the case of a point-mass lens with mass $M_{L}$, 
the bending angle vector $\hat{\balpha}(\bxi)$ is given
by~\citep{SEF}
\begin{equation}
 \hat{\balpha}(\bxi) = 4 M_L \frac{\bxi}{|\bxi|^2}, 
 \label{eq:defan}
\end{equation}
where $\bxi$ is the impact vector (see Fig.~\ref{fig:lens}). 
For convenience, we consider the ``straight" line $A$ 
from a source to the observer and define the intersection point 
of $A$ with the lens plane as the origin of the lens position 
$\bzeta$ and the ray position $\bgamma$. 
We note that, in general, the vector $\bxi$ is 
not orthogonal to the line $A$. However, if the geometrically thin lens 
approximation is valid, we can regard the vectors $\bxi$, $\bzeta$, 
and $\bgamma$ as parallel to each other~\citep{SEF}. 

We define $D_S$, $D_L$, and $D_{LS}$ as the angular diameter 
distances from the observer to
the source, from the observer to the lens, and from the lens to the
source, respectively.
Using simple trigonometry~(see Fig.~\ref{fig:lens}),
we find the relation between the source position $\boeta$ and
$\bxi$, 
\begin{equation}
 \boeta
 = \frac{D_S}{D_L} \bxi - D_{LS} \hat{\balpha}(\bxi).
 \label{eq:lens}
\end{equation}
%
Equation~(\ref{eq:lens}) becomes
\begin{equation}
 0
 = \frac{D_S}{D_L} \bgamma
   - D_{LS} \hat{\balpha}(\bgamma - \bzeta)
\label{eq:lens2}
\end{equation}
in terms of $\bzeta$ and $\bgamma$.

Here we introduce the dimensionless vectors defined by
\begin{eqnarray}
\bx=\frac{\bxi}{\xi_0}~~~~~{\rm and}~~~~~ 
\bl=\frac{\boeta}{\eta_0}, 
\end{eqnarray}
where $\xi_0$ is the Einstein radius given by 
\begin{equation}
 \xi_0 = \sqrt{4 M_L \frac{D_L D_{LS}}{D_S}} \label{eq:re}
\end{equation}
and
\begin{equation}
 \eta_0 = \frac{D_S}{D_L} \xi_0. 
\end{equation}
From Figure~\ref{fig:lens}, we find  the relation 
\begin{equation}
 \bl = - \frac{\bzeta}{\xi_0}. 
\end{equation}
The source will have two images at position 
$(\bx_+, \bx_-)$. 
Substituting equation~(\ref{eq:defan}) into equation~(\ref{eq:lens})
we find 
\begin{equation}
 x_\pm = \frac{1}{2} \left(\sqrt{l^2 + 4} \pm l \right)
 \label{eq:lens3}, 
\end{equation}
where 
$x_\pm = |\bx_\pm| $ and 
$l = |\bl| $. 
The total magnification factor $\mu$ is given by 
\begin{equation}
\mu=\left|\mu_+\right|+\left|\mu_-\right|
=\left|\det\left(\frac{\partial \bl}
{\partial \bx_+}\right)\right|
+\left|\det\left(\frac{\partial \bl}
{\partial \bx_-}\right)\right|=\frac{l^2+2}{l\sqrt{l^2+4}}. 
\label{eq:totmag}
\end{equation}

\subsection{Wave optics in gravitational lensing due to point mass lens}
\label{sec:wave}
For simplicity, let us consider gravitational waves that propagate 
in asymptotically flat spacetime with the metric 
\begin{equation}
 ds^2 = -(1 + 2U) dt^2 + (1 - 2U) d\br^2,
 \label{Newtonian metric}
\end{equation}
where $U$ is the gravitational potential and we assume $|U|\ll1$.  
If the wavelength of the gravitational waves is much smaller than
the typical curvature radius of the background spacetime,
the equation for the gravitational waves is equivalent to that of scalar
waves.
Assuming monochromatic waves of an angular frequency $\omega$,
we have~\citep{Peters:1974}
\begin{equation}
 \left(\nabla^2 + \omega^2 \right) \phi = 4 \omega^2 U \phi. 
 \label{eq:wave2}
\end{equation}

For convenience, we introduce the amplification factor $F$ defined by  
\begin{equation}
 F = \frac{\phi^L(\omega)}{\phi(\omega)}, 
\end{equation}
where $\phi$ is the plane wave with no lensing effects, 
and $\phi^L$ is the wave that is undergoing lensing effects.
Here note that $\phi^L$ is no longer a plane wave,
although it was initially.
In the case of the point-mass lens with mass $M_{L}$,
the gravitational potential is given by $U=-M_{L} / r$. 
Then we have the solution of equation~(\ref{eq:wave2}) in the form 
\begin{equation}
 F(\omega;y)
 = \exp\left[
         \frac{\pi w}{4} + i \frac{w}{2} \ln \left(\frac{w}{2}\right)
       \right]
   \Gamma\left(1 - \frac{i}{2} w \right)
   ~_1F_1\left(\frac{i}{2} w, 1; \frac{i}{2} w y^2 \right), 
\label{eq:ampp}
\end{equation}
where $\Gamma$ and $~_1F_1$ are the gamma function and the confluent
hypergeometric function, respectively, 
and $w := 4\omega M_L$~\citep{Peters:1974}.
Since $\left|\nabla U\right|$ is much larger than the Hubble parameter 
of the universe, by replacing $\omega$ with $\omega(1+z_L)$,
we can use equation~(\ref{eq:ampp}) in cosmological
situations.

\subsection{The Dyer-Roeder distance}
\label{sec:adis}

In the clumpy universe model, the local geometry is inhomogeneous, 
but its global aspect might be well described by the FL universe 
whose metric is given by 
\begin{equation}
 ds^2
 = - dT^2
   + a^2(T) \left(\frac{dR^2}{1 + K R^2} + R^2 d\Omega^2 \right), 
\label{eq:homet}
\end{equation}
where $K=1$, $0$, and $-1$, and $d\Omega^2$ is the round metric. 
However, the global aspect of the clumpy 
universe is a vague notion since we do not have a definite mathematical 
prescription to identify the global aspect of the locally 
inhomogeneous universe with the FL universe model. 
Therefore, the above FL universe 
is a fictitious background universe; hereafter, we will refer to this 
fictitious geometry (universe) as the ``global geometry" (or universe). 
We assume that the global geometry is well approximated by 
that of the FL universe 
filled with dust and with a cosmological constant. 

In the clumpy universe model, the distances $D_L$, $D_S$, and $D_{LS}$ 
in equation~(\ref{eq:lens}) or equation~(\ref{eq:lens2}) are 
replaced by the Dyer-Roeder (DR) distances~\citep{Kantowski:1969,Dyer1973}, which are the observed 
angular diameter distances if the gravitational waves propagate far from 
the point masses. The DR distance with smoothness parameter 
$\tilde\alpha=0$ \citep{SEF}
in the global universe defined above 
is given by 
\begin{equation}
 D_{DR}(z_1, z_2)
 = \frac{1 + z_1}{H_0}
   \int^{z_2}_{z_1}
     \frac{1}{(1 + z)^2}
     \frac{dz}{\sqrt{%
                \Omega_{\rm m0} (1 + z)^3-\Omega_{\rm K0} (1 + z)^2
                + \Omega_{\Lambda0}
              }},
\end{equation}
where $z_1$ and $z_2$ are the redshift of the observer and source, 
and $\Omega_{\rm m0}$, $H_0$, and $\Omega_{\rm \Lambda0}$ are  
the present values of the total density parameter, 
Hubble parameter and normalized cosmological constant in the global 
universe, respectively, 
and $\Omega_{\rm K0}=\Omega_{\rm m0}+\Omega_{\Lambda0}-1$. 
Hereafter, the subscript $0$ means the present value.
Some further discussions of the properties 
of the DR distance are presented in \citet{Linder:1988}, \citet{Seitz:1994a} ,
\citet{Kantowski:1998}, and \citet{Sereno:2001}. 
\section{The distribution of point masses}
\label{sec:dispro}

In order for the clumpy universe to be the same as a homogeneous 
and isotropic universe from a global perspective, the point masses must 
be distributed uniformly. However, since there are no isometries of 
homogeneity and isotropy in the clumpy universe, 
the notion of ``uniform distribution" cannot be introduced in a strict 
sense. Thus, we assume a uniform distribution 
of the point masses with respect to the global geometry 
represented by equation~(\ref{eq:homet}) 
in a manner consistent with the mass density of this global 
universe. 
This universe model is called ``the on-average Friedmann universe." 
The assumptions in this universe model are discussed in \citet{Seitz:1994b}. 

The comoving number density $\rho_{\rm n}$ of the point masses $M_{L}$ 
is given by 
\begin{equation}
 \rho_{\rm n}
 = \frac{a^3 \rho}{M_L}
 = \frac{3 \Omega_{\rm m0} H_0^2}{8 \pi M_L}, 
\end{equation}
where $\rho$ is the average mass density and we have set $a_0=1$. 
We consider a past light cone in the global universe. The vertex of this 
light cone is at the observer, at $R=0$ and $z=0$, 
and is parametrized by the redshift $z$ of its null geodesic generator. 
Then the comoving volume of a spherical shell bounded by $R(z)$ and  
$R=R(z+\Delta z)$ on this light cone is given by 
\begin{equation}
 \Delta V = 4 \pi R^2 \frac{dR}{dz} \Delta z. 
\end{equation}
Therefore, the number of point masses in this shell is given by  
\begin{equation}
 \Delta N
 = \rho_{\rm n} \Delta V
 = \frac{3 \Omega_{\rm m0} H_0^2}{2 M_L} R^2 \frac{dR}{dz} \Delta z, 
 \label{eq:dN}
\end{equation}
where 
\begin{equation}
 \frac{dR}{dz}
 = \frac{1}{H_0}
   \sqrt{\frac{1 + H_0^2 \Omega_{K0} (1 + z)^2 D_F^2(z)}
              {\Omega_{\rm m0}(1 + z)^3 - \Omega_{K0} (1 + z)^2
               + \Omega_{\Lambda0}}
        },
\label{eq:dR}
\end{equation}
where $D_F(z)$ is the angular diameter distance 
from the source of the redshift $z$ to the observer in the 
global universe, which is expressed as 
\begin{equation}
 D_F(z)=\left\{
 \begin{array}{ll}
\displaystyle
  \frac{1}{H_0 (1 + z)}
  \frac{\sin X(z)}
        {\sqrt{\Omega_{\rm K0}}}
  &~~{\rm for}~~
  \Omega_{\rm K0} > 0 \\
\displaystyle  
  \frac{1}{H_0 (1 + z)} Y(z)
  &~~{\rm for}~~
  \Omega_{\rm K0} = 0 \\
\displaystyle  
  \frac{1}{H_0 (1 + z)}
  \frac{\sinh X(z)}
        {\sqrt{- \Omega_{\rm K0}}}
  &~~{\rm for}~~
  \Omega_{\rm K0} < 0 
\end{array}\right., 
\end{equation}
where 

\begin{equation}
 X(z)
 = \sqrt{\left|
            \Omega_{\rm K0}
          \right|}
    \int^{z}_{0}
      \frac{dz'}
           {\sqrt{
             \Omega_{\rm m0} (1 + z')^3 
             -\Omega_{\rm K0}  (1 + z')^2
             + \Omega_{\Lambda0}
           }},
\end{equation}
\begin{equation}
 Y(z)
 = \int^{z}_{0}
      \frac{dz'}
           {\sqrt{
              \Omega_{\rm m0} (1 + z')^3
              - \Omega_{\rm K0} (1 + z')^2
              + \Omega_{\Lambda0}
           }}.
\end{equation}
The average number of point masses in the region
$y < \zeta / \xi_0 < y + \Delta y$ of this shell is given by 
\begin{equation}
 p(y,z) \Delta y
 = \frac{2\pi \xi_0^2 y \Delta y}{4 \pi a^2 R^2} \Delta N.
 \label{eq:p1}
\end{equation}
Substituting equations~(\ref{eq:dN}) and (\ref{eq:dR}) into 
equation (\ref{eq:p1}) and 
using equation~(\ref{eq:re}), we have
\begin{eqnarray}
 p(y,z) \Delta y
 = 3 H_0 \Omega_{\rm m0} (1 + z)^2
   \sqrt{\frac{1 + H_0^2 \Omega_{K0} (1 + z)^2 D_F^2(z)}
              {\Omega_{\rm m0}(1 + z)^3 - \Omega_{K0} (1 + z)^2
               + \Omega_{\Lambda0}}}
 \nonumber&&\\
 \hspace{7cm}\times
 \frac{D_{DR}(0, z) D_{DR}(z, z_S)}{D_{DR}(0, z_S)}
 y \Delta y \Delta z,&& 
 \label{eq:p3}
\end{eqnarray}
where $z_S$ is the redshift of the source and we have replaced 
each of distances 
in equation~(\ref{eq:re}) with the DR distance.  
We find that $p(y, z)$ does not depend on $M_L$, and therefore  
the lensing probability for a given source does not depend on $M_L$. 

We note that the assumption of the distribution of 
point masses given in equation (\ref{eq:p3}) does not necessarily guarantee 
equality between the magnification in the 
global universe defined by 
\begin{equation}
\mu_F:=\left(\frac{D_{DR}(0,z)}{D_F(z)}\right)^2
\label{eq:mufl}
\end{equation}
and the average magnification in the 
clumpy universe. Sevral previous works have discussed this issue
~\citep{Weinberg:1976,Ellis:1998ha,Claudel:2000ug,Rose:2001qi,Kibble:2004tm}. 

\section{The long-wavelength case, $\lambda\gg M_L$}
\label{sec:wlim}
\begin{figure}[htbp]
\begin{center}
\includegraphics[scale=0.8]{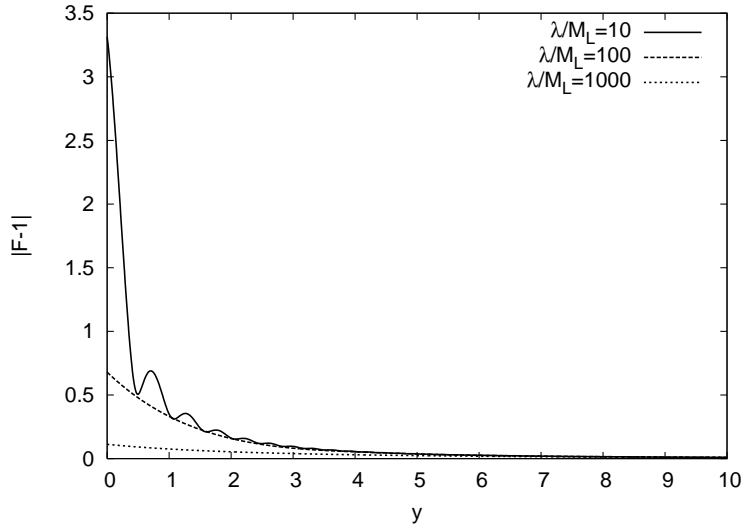}
\caption{Amplification factor $F(\omega(1+z_{L});y)$ given in
eq.~(\ref{eq:ampp}) with replacement
$\omega \rightarrow (1 + z_{L}) \omega$.
We plot $\left|F-1\right|$ as a function of $y$ for three cases of
$\lambda / M_L=10, 10^2$, and $10^3$. 
The redshift of the source $z_{S}$ and of the lens $z_{L}$ are fixed for 
$z_{S} = 2$ and $z_{L} = 1$.
}
\label{fig:amp}
\end{center}
\end{figure}

In the case of $\lambda \gg M_L$, 
the amplification factor $F(\omega(1+z_L);y)$ is almost equal to 
unity~(see Fig.~\ref{fig:amp}) because of diffraction 
effects~\citep{Nakamura:1997sw}, and we have 
\begin{equation}
 \frac{|\phi_L-\phi|}{|\phi|}
  =|F(\omega(1+z_L);y)-1|\ll1.
\label{eq:ll1}
\end{equation}
Thus we assume that a wave $\phi$ can be treated as a spherical wave 
whose center is at the source during all stages of the propagation.
Let us consider the situation in which $N$ point masses are located at
$(z_i, y_i),~i=1,\dots, N$,, 
where $i$ is the number assigned to each point mass in order of distance
from the observer. 
If our assumption is valid, the wave $\phi$ is changed whenever it propagates near each point
mass as 
\begin{equation}
 \phi\rightarrow F_N\phi \rightarrow F_NF_{N-1}\phi 
 \rightarrow \cdots \rightarrow \prod_{i=1}^NF_i~\phi, 
\end{equation}
where 
\begin{equation}
F_i=F\left(\omega\left(1+z_i\right);y_i\right). 
\end{equation}
Thus the observed amplitude $|\phi_{\rm obs}|$ is given by
\begin{equation}
 |\phi_{\rm obs}|
 = \prod_{i=1}^N\left|F_i\right|\left|\phi\right|. 
\end{equation}
Then, the total amplification $\left| F_{\rm total}\right|$ is given by 
\begin{equation}
 \left|F_{\rm total}\right|
 = \prod_{i=1}^N\left|F_i\right|.
\end{equation}

\begin{figure}[htbp]
\begin{center}
\includegraphics[scale=0.8]{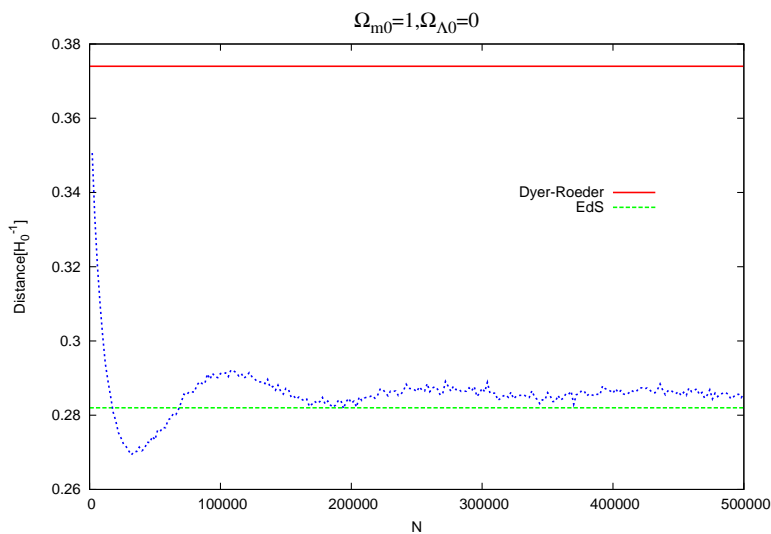}
\caption{Observed distance $D_{\rm obs}$, depicted as 
a function of $N$ in the case of $\lambda/M_L=10^6$. 
The solid line represents the DR distance in the EdS universe. 
The dashed line represents the angluar diameter distance in 
the global universe. 
}
\label{fig:forNeds}
\end{center}
\end{figure}

\begin{figure}[htbp]
\begin{center}
\includegraphics[scale=0.8]{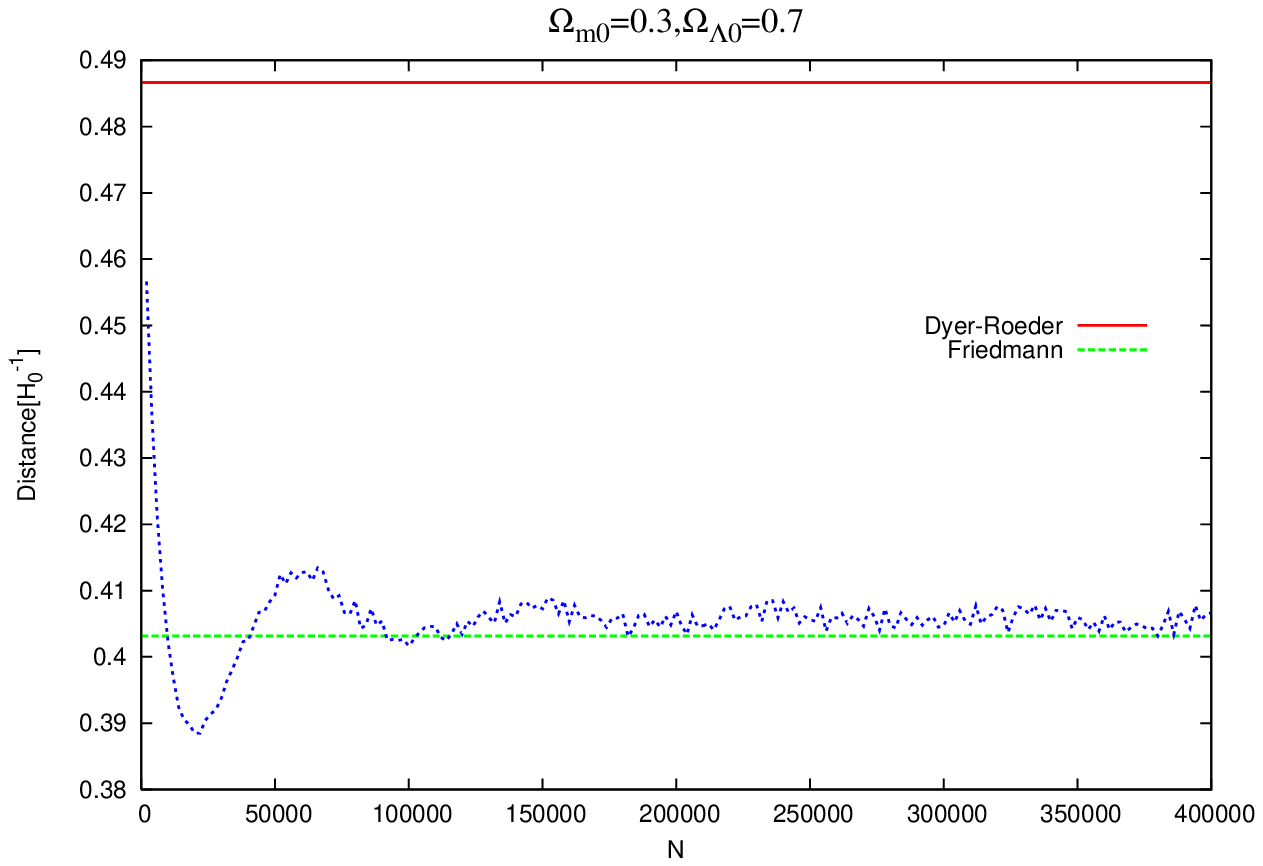}
\caption{Same as in Fig.~\ref{fig:forNeds}, but here the global 
geometry is given by the FL universe with 
$\Omega_{\lambda 0}=0.7$ and $\Omega_{\rm m0}=0.3$.
}
\label{fig:forNlcdm}
\end{center}
\end{figure}

We randomly distribute point masses so that the distribution
of those is consistent with equation~(\ref{eq:p3}).
First, we divide the spherical region $z<z_S$ into $N$ concentric
spherical shells, each of which is bounded by two spheres, 
$z = z_i - \Delta z / 2$ and $z = z_i + \Delta z / 2$.
We take into account only the nearest lens in each shell.
As described below, this procedure is valid for large numbers $N$, 
i.e., small $\Delta z$. 
We find from equation~(\ref{eq:p3}) that in the $i$th shell
$z_i - \Delta z / 2 < z < z_i + \Delta z / 2$, there is one point mass 
within the region $y\leq Y_i$ on average, where $Y_i$ is defined by 
\begin{equation}
 \int^{Y_i}_0 p(y, z_i) dy=1.
\end{equation}
Then we randomly put a point mass within the region $y\leq Y_i$ in the
$i$th shell.
Here it is worth noting that $Y_i$ is an increasing 
function of $N$, so by setting $N$ large, we 
can take into account lenses far from the line $A$. 
Hence, in the limit of $N\rightarrow\infty$, we can all lenses take 
into account. 
The results do not depend on the value of $N$, 
if it is large enough~
(see Figs.~\ref{fig:forNeds} and \ref{fig:forNlcdm}). 
In our calculations, we set $N$ sufficiently large. 

\begin{figure}[htbp]
 \begin{center}
  \includegraphics[scale=0.7]{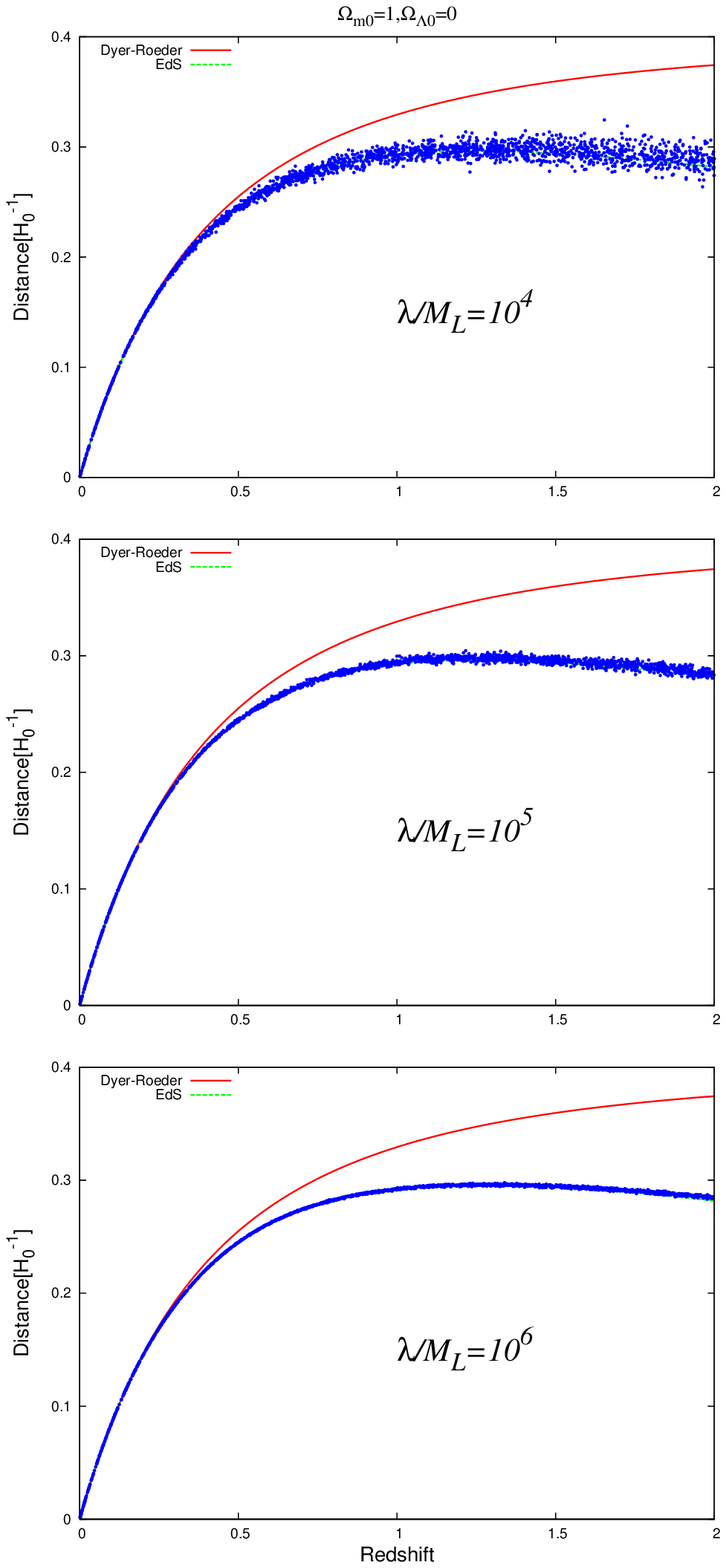}
  \caption{Distance-redshift relation of 2000 samples, 
  depicted
  in the long-wavelength cases $\lambda / M_{L} \gg 1$. 
  We have set $N=1,000+\lfloor 10,000\times z_S \rfloor$ 
  in the cases of $\lambda / M_{L} = 10^4$ and $10^5$, and 
  $N=1,000+\lfloor 80,000\times z_S \rfloor$ 
  in the case of $\lambda / M_{L} = 10^6$. 
  We have chosen the EdS universe,
  i.e., the FL model with $\Omega_{\Lambda0}=0$ and $\Omega_{\rm m0}=1$, 
  as a global universe.
  The solid line represents the DR distance in the EdS
  universe. The dashed line which is hidden behind the data points 
  represents the angular diameter distance in
  the global universe. }
  \label{fig:waveeds}
 \end{center}
\end{figure}

\begin{figure}[htbp]
 \begin{center}
  \includegraphics[scale=0.7]{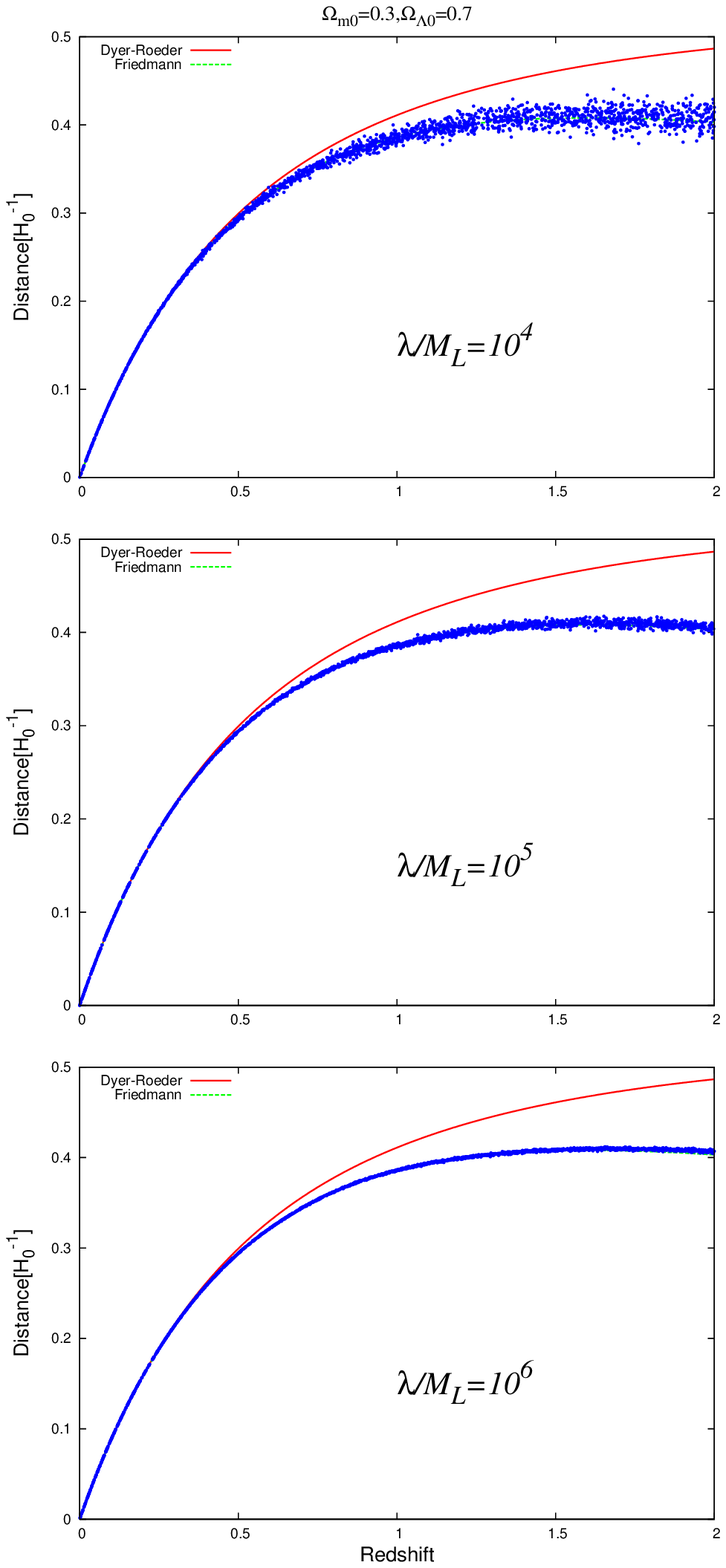}
  \caption{Same as in Fig.~\ref{fig:waveeds}, but here the global 
  geometry is given by the FL universe with $\Omega_{\Lambda0} = 0.7$ 
  and $\Omega_{\rm m0} = 0.3$.}
\label{fig:wavelcdm}
\end{center}
\end{figure}

Calculating the total amplification factor $F_{\rm total}$, 
the observed distance is given by 
\begin{equation}
 D_{\rm obs} = \frac{D_{DR}(0,z_S)}{ |F_{\rm total}|}. 
\end{equation}
We have calculated $2000$ samples in two clumpy universe models, 
the globally Einstein-de Sitter (EdS) and globally FL universe with 
$\Omega_{\Lambda0} = 0.7$ and $\Omega_{\rm m0} = 0.3$. 
The results are depicted in Figures~\ref{fig:waveeds} and
\ref{fig:wavelcdm}, where we have randomly chosen the sources of the 
redshifts in the range $0< z_S\leq2$. 


\begin{figure}[htbp]
\begin{center}
\includegraphics[scale=0.8]{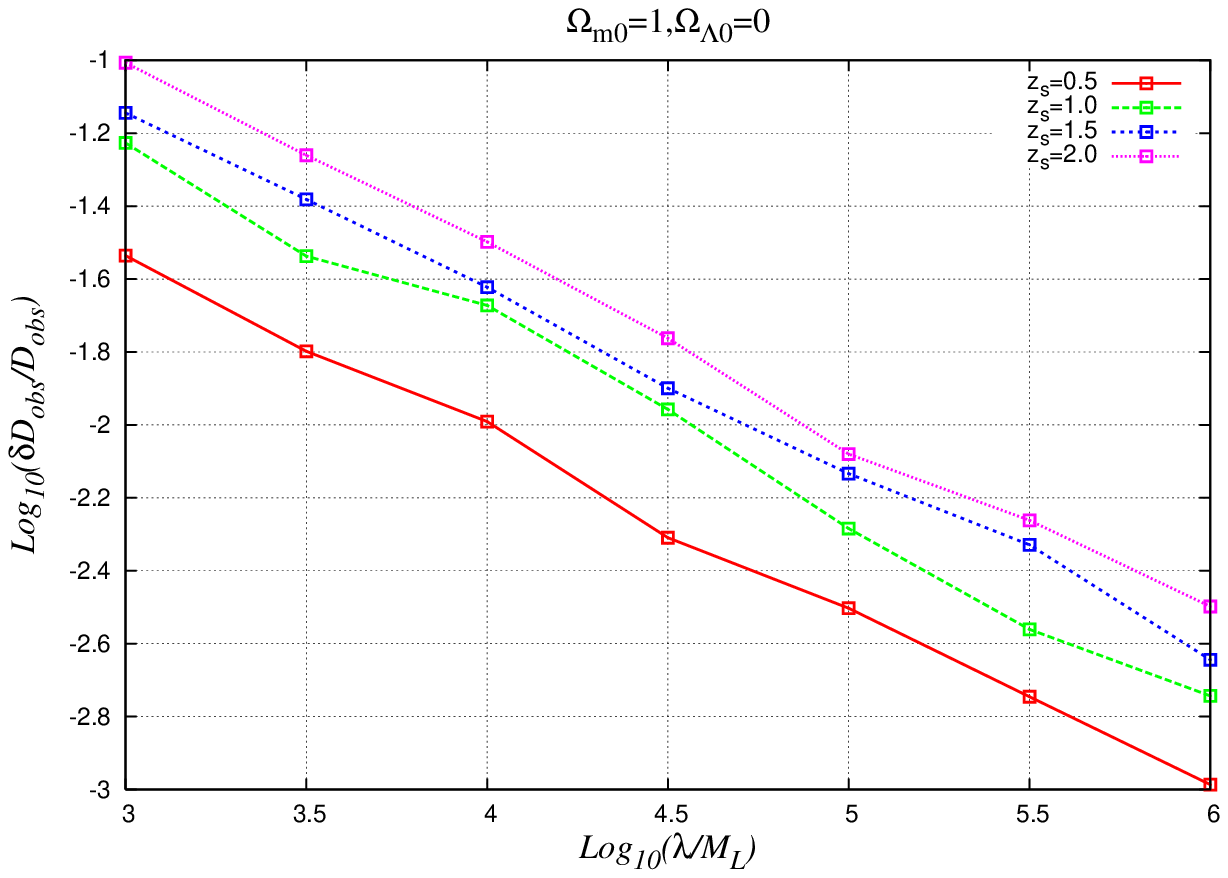}
\caption{Relation between $\delta D_{\rm obs}/\left<D_{\rm obs}\right>$ 
and $\lambda/M_L$, where $\delta D_{\rm obs}$ and 
$\left<D_{\rm obs}\right>$ are the rms deviation and the average 
distance given by 100 samples of $D_{\rm obs}$ in each $\lambda/M_L$ and $z_S$. }
\label{fig:avdeveds}
\end{center}
\end{figure}

\begin{figure}[htbp]
\begin{center}
\includegraphics[scale=0.8]{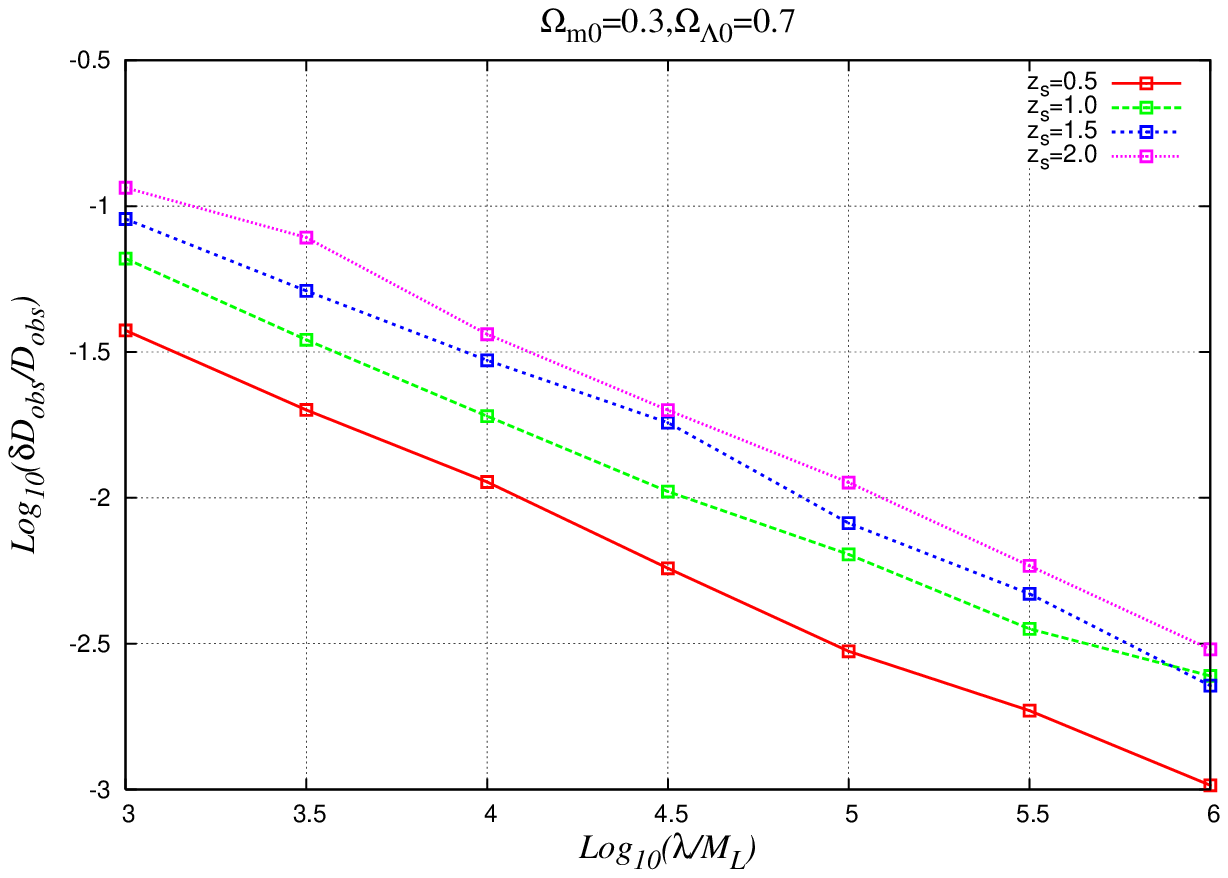}
\caption{Same as in Fig.~\ref{fig:avdeveds}, but here the global 
  geometry is given by the FL universe with $\Omega_{\Lambda0} = 0.7$ 
  and $\Omega_{\rm m0} = 0.3$.
}
\label{fig:avdevlcdm}
\end{center}
\end{figure}

The distance dispersion tends to vanish as $\lambda/M_L$
increases, and $D_{\rm obs}$ is almost equal to $D_F(z_S)$ 
for sufficiently large $\lambda / M_L$~
(see Figs.~\ref{fig:avdeveds} and \ref{fig:avdevlcdm}). 
As described in the next section, in the case of
$\lambda/M_L\ll1$, the dispersion is much larger than for
$\lambda/M_L\gg1$. 
It can be seen from Figures~\ref{fig:avdeveds} and \ref{fig:avdevlcdm} that 
the distance dispersion does not sensitively depend on the value of 
$\Omega_{\rm m0}$, but become large as $z$ increase. 

\section{The short-wavelength case, $\lambda\ll M_L$}
\label{sec:glim}
In the case of $\lambda \ll M_L$,
the value of the amplification factor $F(\omega(1+z_L);y)$ is much
larger than unity for small values of $y$. 
Thus, we cannot use the linear approximation used in the previous
section. Instead, here we adopt the geometrical optics approximation. 
We comment that it is difficult to resolve multiple images in the sky 
using gravitational wave detectors, 
because their angular resolution is 
$\sim 1^\prime$-$1^\circ$, 
much larger than the typical image separation.
In geometrical optics, lensing does not change the wave form 
except for its
 amplitude (due to the magnification) and 
arrival time (due to the time delay) in each image. 
The observed waveform that is given by the 
superposition of waves from each image has 
undergone amplification and phase shift. 
In this paper, we assume stationary configurations, 
and the phase shift is not an observational variable. 
Therefore, we can observe only the amplitude 
of the superposed waves, and we cannot identify the signal 
as a superposed one. 
%
\subsection{The method of calculation}
In order to calculate the amplification factor,
we use the multiple lens-\hspace{0pt}plane method~\citep{SEF}.
We consider the ``straight'' line $A$ from a source to the observer, 
and we put $N$ lens planes between them so that the straight line $A$
intersects these lens planes vertically. 
The results do not depend on the value of $N$ 
if it is large enough; 
we set $N=10^5$ in our calculations.

In the same manner as in \S \ref{sec:wlim},
we determine the redshifts $z_i$ and 
randomly situate a point mass $M_{L}$ within the region $y\leq Y_i$ 
on each lens plane $\Sigma_{i}$.
In the case of $\lambda \gg M_L$,
we have considered the lensing effects due to all point masses 
that we situate within the region $y\leq Y_i$. 
By contrast, in the case of $\lambda\ll M_L$,
we can neglect the lensing effects due to the point masses that 
are far from the ray. 
Thus we take into account lensing effects due to the point masses
in the region $y < y_{\rm max}$, where $y_{\rm max}$ is some positive 
constant smaller than $Y_i$. 
In other word, we take into account only a few lens planes 
in which a point mass is put within the region $y\leq y_{\rm max}$. 
Hereafter we denote the number of lens planes that 
we take into account as $N_L$. 
The validity of this procedure will be supported by the discussion 
of methods for determining  an appropriate value of $y_{\rm max}$ 
in \S \ref{sec:dnp}. 

We denote the lens planes from the observer to the source 
as $\Sigma_1, \Sigma_2, \dots, \Sigma_{N_L}$.
For convenience, we name the source plane $\Sigma_{N_L + 1}$, 
which is also orthogonal to the line $A$. 
We define the intersection point of $A$ with 
$\Sigma_i$ as the origin of the lens position $\bzeta_i$ and 
the ray position $\bgamma_i$. 
Suppose lens positions $\bzeta_i~(i = 1, \dots, N_L)$ are given.
Then once the ray position $\bgamma_1$ is given on the first lens plane $\Sigma_1$, 
the ray position $\bgamma_j$ 
is recursively given by 
\begin{equation}
 \bgamma_j
 =   \frac{D_j}{D_1} \bgamma_1
   - \sum_{i\leq j-1} D_{ij} \hat{\balpha}(\bgamma_i - \bzeta_i) 
 \label{eq:tikuji} 
\end{equation}
on the 
$j$th lens plane $\Sigma_j$ $(j \leq N_L + 1)$, 
where $D_1=D_{DR}(0,z_1)$, $D_j=D_{DR}(0,z_j)$, and $D_{ij}=D_{DR}(z_i,z_j)$. 
For later convenience,
we specify the ray positions and the lens positions by the
following dimensionless quantities:
\begin{eqnarray}
 \bu_i &:=& \frac{\bgamma_i}{D_i},\\
 \bq_i &:=& \frac{\bzeta_i}{D_i}.
\end{eqnarray}
Then equation (\ref{eq:tikuji}) becomes 
\begin{equation}
 \bu_j
 =   \bu_1
   - \sum_{i\leq j-1} \beta_{ij} \balpha(\bu_i - \bq_i),
   \label{eq:tikuji2}
\end{equation}
where denoting $D_{1N_L+1}$ and $D_{iN_L+1}$ by $D_S$ and $D_{iS}$, respectively, 
\begin{equation}
 \beta_{ij} := \frac{D_{ij} D_S}{D_j D_{iS}}
\end{equation}
and
\begin{equation}
 \balpha(\bu_i - \bq_i)
 := \frac{D_{iS}}{D_S} \hat{\balpha}(\bgamma_i - \bzeta_i).
\end{equation}
\begin{figure}[htbp]
\begin{center}
\includegraphics[scale=0.6]{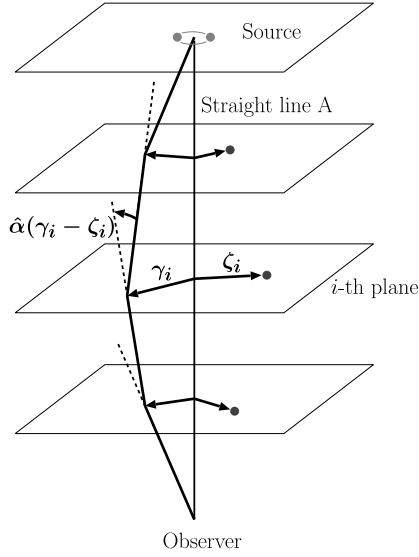}
\caption{Definitions of the vectors $\balpha_j$, $\bzeta_j$ and
 $\bgamma_j$. 
}
\label{fig:multi}
\end{center}
\end{figure}

Since more than one image appears in the present case, 
the amplification factor $F_{\rm total}$ is determined 
by both the interference  
of the rays corresponding to those images and 
the magnification of each ray due to the gravitational focusing 
effects. We specify a ray from the source to the observer by the 
ray position $\bu_1=\bov_{(p)}$ on the first lens plane $\Sigma_1$, where 
we have introduced the label $p$ to distinguish between the rays. 
The magnification $\mu_{(p)}$ for the ray $p$ is defined by 
\begin{equation}
 \mu_{(p)} = \frac{1}{\det \mathcal{A}_{(p)}},
 \label{eq:mup}
\end{equation}
where $\mathcal{A}_{(p)}$ is the Jacobian matrix of the lens mapping,
\begin{eqnarray}
 \mathcal{A}_{(p)}
 &=& \left.
      \frac{\partial \bu_{N + 1}}{\partial \bu_{1}}
    \right|_{\bu_1 = \bov_{(p)}} \nonumber \\
 &=& \mathcal{I}
      - \sum^{N}_{i=1} \beta_{i, N+1}
          \left.
            \frac{\partial \balpha(\bu_i - \bq_i)}{\partial \bu_i}
            \frac{\partial \bu_i}{\partial \bu_1}
          \right|_{\bu_1 = \bov_{(p)}}.
 \label{eq:jaco}
\end{eqnarray}
The amplification factor $F_{\rm total}$ is then given by 
\begin{equation}
 F_{\rm total}
 = \sum_p |\mu_{(p)}|^{1/2} \exp (i \theta_{(p)}),
\label{eq:geoamp}
\end{equation}
where the phase $\theta_{(p)}$ is determined by the time delay
$T_{(p)}$ relative to the ray that does not receive any lensing
effect, and by the number of caustics $n_{(p)}$ on ray $p$~\citep{CTF}, 
in the manner 
\begin{equation}
 \theta_{(p)}
 := \omega T_{(p)} - \frac{\pi}{2} n_{(p)}.
\end{equation}
The time delay $T_{(p)}$ is given by~\citep{SEF}
\begin{equation}
 T_{(p)}
 = \left. \sum_{i=1}^{N_L}
            (1 + z_i)
            \left\{
              \frac{D_i D_{i+1}}{2 D_{i,i+1}} (\bu_i - \bu_{i+1})^2
              - 4M_L \ln \left| \bu_i - \bq_i \right|
            \right\}
   \right|_{\bu_1 = \bov_{(p)}},
 \label{eq:timeij}
\end{equation}
and the number of caustics $n_{(p)}$ is give by 
\begin{equation}
 n_{(p)} = \sum_{i=1}^{N_L-1} n_{i(p)}, 
\end{equation}
where $n_{i (p)}$ is the number of caustics between 
$\Sigma_{i}$ and $\Sigma_{i + 1}$ on ray $p$ and its value is 
given by 
\begin{equation}
 n_{i(p)} = \left\{
 \begin{array}{l}
\displaystyle
  0 ~~{\rm for}~~
    \det \left(
           \frac{\partial \bu_{i+1}}{\partial \bu_i}
         \right) > 0
    ~~{\rm and}~~
    \tr \left(
          \frac{\partial \bu_{i+1}}{\partial \bu_i}
        \right) > 0 \\
\displaystyle  1 ~~{\rm for}~~
    \det \left(
          \frac{\partial \bu_{i+1}}{\partial \bu_i}
         \right) < 0  \rule{0pt}{30pt}\\
\displaystyle  2 ~~{\rm for}~~
    \det \left(
           \frac{\partial \bu_{i+1}}{\partial \bu_i}
         \right) > 0
    ~~{\rm and}~~
    \tr \left(
          \frac{\partial \bu_{i+1}}{\partial \bu_i}
        \right) < 0 \rule{0pt}{30pt}
 \end{array}\right. .
\end{equation}
The square of the absolute value of $F_{\rm total}$ is given by 
\begin{equation}
 |F_{\rm total}|^2
 =   \sum_p |\mu_{(p)}|
   + \sum_{p \neq q}
       |\mu_{(p)}|^{1/2} |\mu_{(q)}|^{1/2}
       \cos \left(\theta_{(p)} - \theta_{(q)}\right).
 \label{eq:ampmag}
\end{equation}
The first term on the right-hand side of equation (\ref{eq:ampmag})
represents the shear and focusing effect on the ray bundle, and the second term
represents the interference effect~\citep{Nakamura:1999}.

Finally, we explain the method for finding rays from the source to the
observer. If the lens positions $\bzeta_i$ ($i=1,. . .,N_L$) are given, 
the dimensionless ray position $\bu_{N_L+1}$ on the source plane
$\Sigma_{N_L+1}$ is completely 
determined by the ray position $\bu_{1}$ on the
first lens plane $\Sigma_{1}$;
\begin{equation}
 \bu_{N_L+1} = \bu_{N_L+1}(\bu_{1}). \label{eq:lens mapping}
\end{equation}
Therefore, $\bov_{(p)}$ is a root of the equation
\begin{equation}
 \bu_{N_L+1}(\bov_{(p)}) = 0. \label{eq:map0}
\end{equation}
We solve equation (\ref{eq:map0}) using
the Newton-\hspace{0pt}Raphson method,
which requires initial estimates for the desired roots~\citep{rauch}.
In order to find appropriate initial estimates,
we consider a square region on the first lens plane $\Sigma_{1}$
whose center agrees with the intersection point of the straight line
$A$ with $\Sigma_1$~(see Fig.~\ref{fig:multi}) 
and with sides of length 
$2y_{\rm max}\xi_{01}$, where $\xi_{01} = \sqrt{4M_L D_1 D_{1S} / D_S}$.
In this square region, 
we place evenly spaced $n\times n$ grid points whose intervals are equal
to $2 y_{\rm max} \xi_{01} / n$.
Substituting the position vector of each grid point to
$\bu_1$ on the right hand side of equation (\ref{eq:lens mapping}), 
we obtain the corresponding ray position $\bu_{N_L+1}$ on the source
plane.
Then, we adopt the grid points which lead to 
\begin{equation}
 \frac{\gamma_{N_L+1}}{\sqrt{M_L D_S}}
 = \frac{D_S u_{N_L+1}}{\sqrt{M_L D_S}} < 1
 \label{eq:mato}
\end{equation}
as the initial estimates of the Newton-Raphson method.

In practice, the interval between the grid points 
must be finite for the numerical calculations, 
so we may fail to find initial estimates that lead to some solutions
$\bov_{(p)}$.
However, the magnification factors $\left|\mu_{(p)}\right|$ of such
solutions are so small that the contributions to
$\left|F_{\rm total}\right|$ will be negligible.
In order to see this,
let us consider the inverse map of equation (\ref{eq:lens mapping}).
The inverse image of $S$ on the first lens plane consists of more than one 
connected region since multiple images of the source appear due to
the lensing effects.
If a grid point is contained in one of the connected regions,
the grid point is found as an initial estimate by imposing the above
criterion given by (\ref{eq:mato}).
If there is no grid point in a connected region due to its smallness,
we cannot find initial estimates for the solution $\bov_{(p)}$ included
in the connected region and cannot find the solution $\bov_{(p)}$ itself.
Here we should note that the magnification factor associated with a root
obtained from initial estimates contained in a connected region is
almost proportional to the area of the connected region divided by the
area of $S$.
Therefore the magnification factor associated with the root we have failed to
find will be small.
For every initial estimate, we obtain the roots $\bov_{(p)}$.
However, these roots do not necessarily differ; 
the different initial estimates may lead to an identical root.
Therefore, in order to prevent multiple counting, we have to take
paths that differ from each other.

\subsection{Determination of numerical parameters} 
\label{sec:dnp}
We determine appropriate numerical values of $n$ and $y_{\rm max}$ 
to find all solutions $\bov_{(p)}$ of equation (\ref{eq:map0})
whose magnification factors $\left|\mu_{(p)}\right|$ significantly 
contribute to $\left|F_{\rm total}\right|$. 
If we fail to find such solutions, we underestimate 
the total flux that emanates from the source to the observer. 
Hence, this is a very important task. 
Since the interference effect is not important for our present purpose, 
we neglect it for simplicity. Then the total flux is proportional to 
the magnification $\mu$, defined by 
\begin{equation}
\mu=\sum_p\left|\mu_{(p)}\right|. 
\end{equation}
If we 
increase the values of $n$ and $y_{\rm max}$, 
the magnification $\mu$ will approach the maximum value $\mu_{\rm max}$. 
We should determine the values of $n$ and $y_{\rm max}$ 
so that the magnification $\mu$ satisfies $\mu\simeq\mu_{\rm max}$. 
Since the lens distribution is determined using  
pseudorandom numbers, the magnification $\mu$ 
necessarily has a stochastic nature, and therefore we have to consider the 
ensemble average of $\mu$. 

First, we set the source redshift $z_S$ to be $2$ and numerically generate samples of 
lens distributions with numbers much larger than 20,000. 
Then, with the parameters $y_{\rm max}$ and $n$ fixed, 
we calculate lens mappings 
to obtain the magnification $\mu$ for each lens distribution 
and construct an ensemble composed of 20,000 samples of $\mu$. 
This procedure is repeated to construct ensembles 
for various values of parameters $y_{\rm max}$ and $n$.

We denote one of the ensembles by $\cal{E}$, and 
further denote the ensemble average of a quantity $Q(\mu)$ related the magnification $\mu$ 
in $\cal{E}$ by $\left<Q\right>_{\cal{E}}$. 
The reader might think that $\left<\mu\right>_{\cal E}$ will approach $\mu_{\rm max}$ in the 
limit of large $y_{\rm max}$ and $n$, and thus we might be able to choose the parameters $y_{\rm max}$ and 
$n$ so that $\left<\mu\right>_{\cal E}$ sufficiently converges to some value 
that will be 
equal to $\mu_{\rm max}$. However this is not the case, 
for the following reason. 
As shown in \S \ref{sec:nr}, our numerical simulations imply that 
the probability $P(\mu)d\mu$ that the 
received magnification is in the range $(\mu,\mu+d\mu)$ 
is proportional to $\mu^{-3}$ for sufficiently large $\mu$. 
Therefore the second moment of the magnification $\left<\mu^2\right>$, defined by
\begin{equation}
\left<\mu^2\right>\equiv \int_0^\infty \mu^2 P(\mu)d\mu
\end{equation}
will diverge. 
This implies that the law of large numbers fails, 
or in other words, $\left<\mu\right>_{\cal{E}}$ cannot approximate 
the true average value $\left<\mu\right>$ even if the number of samples 
in the ensemble $\cal{E}$ is much larger than unity
~\citep{Paczynski:1986,Holz:1997ic,rauch}. 
Thus $\left<\mu\right>_{\cal{E}}$ cannot be a criterion for determining 
whether the values of the 
parameters $y_{\rm max}$ and $n$ are valid. 

To obtain the criterion for determining 
validity of the choices of $y_{\rm max}$ and $n$, 
here we set an upper threshold $\mu_{\rm u}$ on $\mu$, 
and replace the magnification of samples in $\cal{E}$ by $\mu_{\rm u}$,  
if those are larger than $\mu_{\rm u}$. 
We denote the ensemble obtained by this procedure as $\hat{\cal E}$. 
The probability distribution of the samples in the ensemble $\hat{\cal E}$ 
will be well approximated by 
\begin{equation}
\hat{P}(\mu)=\theta(\mu_{\rm u}-\mu)P(\mu)+\delta(\mu-\mu_{\rm u})
\int^\infty_{\mu_{\rm u}}P(\bar{\mu})d\bar{\mu}
\end{equation}
if $y_{\rm max}$ and $n$ are large enough, where $\theta(z)$ is a Heaviside's step function, 
and $\delta(z)$ is a Dirac delta function. 
Then we expect
\begin{equation}
\left<\mu\right>_{\hat{\cal E}}\simeq 
\int^\infty_0\mu\hat{P}(\mu)d\mu.
\end{equation}
We use $\left<\mu\right>_{\hat{\cal E}}$ as a criterion 
for the validity of the ensemble $\cal{E}$. 

\begin{figure}[htbp]
\begin{center}
\includegraphics[scale=0.8]{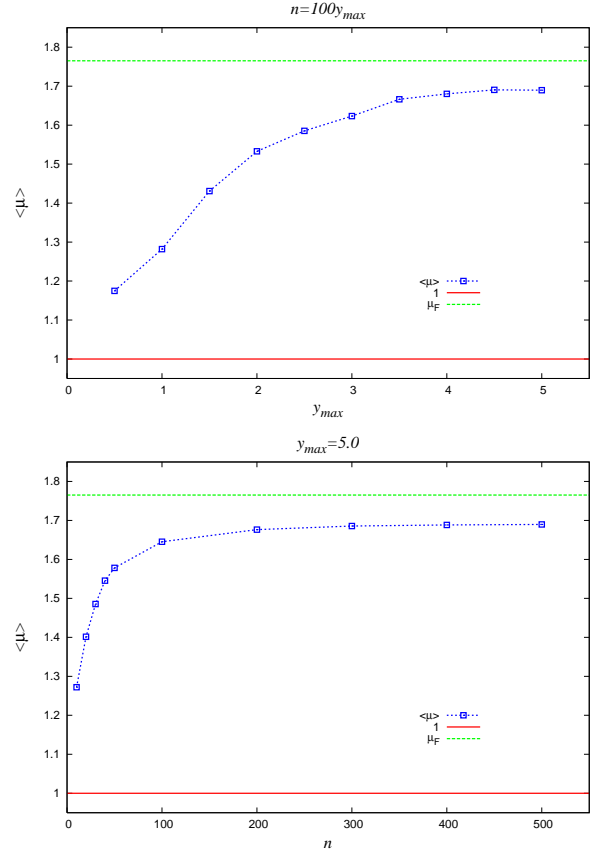}
\caption{Ensemble average of magnification factor 
$\left<\mu\right>_{\hat{\varepsilon}}$, 
depicted as a function of the numerical 
parameters $n$ and $y_{\rm max}$. 
We have set $z_S=2$ and chosen the EdS universe as a global 
universe. 
The dashed horizontal line represent the value of $\mu_F$. 
}
\label{fig:conveds}
\end{center}
\end{figure}

\begin{figure}[htbp]
\begin{center}
\includegraphics[scale=0.8]{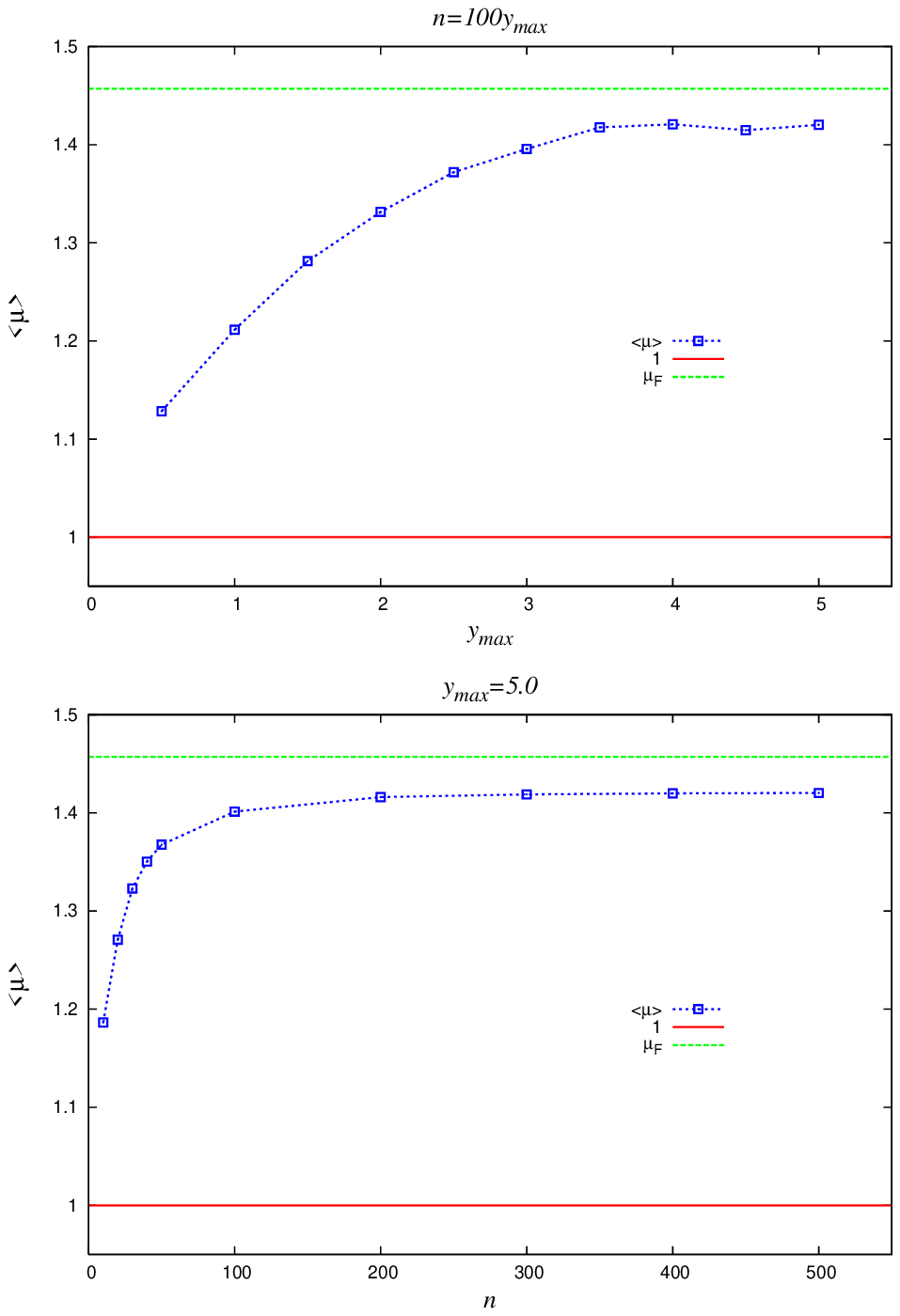}
\caption{Same as in Fig.~\ref{fig:conveds}, 
but here the global geometry is given by the FL universe with 
$\Omega_{\Lambda 0}=0.7,~\Omega_{{\rm m}0}=0.3$. }
\label{fig:convlcdm}
\end{center}
\end{figure}

The relation between $\left<\mu\right>_{\hat{\cal E}}$ and 
the numerical parameters $y_{\rm max}$ and $n$ 
is depicted in Figures~\ref{fig:conveds} and \ref{fig:convlcdm}, where 
we have set $\mu_{\rm u}=100$. The value of $\left<\mu\right>_{\hat{\cal E}}$ 
almost converges to the maximum values 
at $y_{\rm max}=5, n=500$. 
In the case where the numerical parameters $y_{\rm max}$ and $n$ 
are not sufficiently large, there are lens distributions $(z_i,y_i)$ 
in which we cannot find any image. 
We exclude such cases as errors. 
In Table~\ref{tab:error} we list the number of such errors 
appearing as we collected $20,000$ samples. 
Taking these results into account, 
hereafter we set $y_{\rm max}=5$ and $n=500$. 
\begin{table*}[htbp]
\begin{center}
\caption{Number of errors in which we cannot find any image 
of the calculation for Figures 
\ref{fig:conveds} and \ref{fig:convlcdm} is given here. 
\label{tab:error}
}
\begin{tabular}{cc|cc|cc|cc}
\tableline\tableline
\multicolumn{4}{c|}{$\Omega_{\Lambda0}=0,\Omega_{\rm m0}=1$}&
\multicolumn{4}{|c}{$\Omega_{\Lambda0}=0.7,\Omega_{\rm m0}=0.3$}\\
\tableline
\multicolumn{2}{c|}{$n=100\times y_{\rm max}$}&\multicolumn{2}{|c|}{$y_{\rm max}=5$}&
\multicolumn{2}{|c|}{$n=100\times y_{\rm max}$}&\multicolumn{2}{|c}{$y_{\rm max}=5$}\\
\tableline
$y_{\rm max}$&error&$n$&error&$y_{\rm max}$&error&$n$&error\\
\tableline
$0.5$&$12$&$10$&$11791$&$0.5$&$6$&$10$&$8495$\\
$1.0$&$0$&$20$&$1294$&$1.0$&$0$&$20$&$864$\\
$1.5$&$0$&$30$&$290$&$1.5$&$0$&$30$&$168$\\
$2.0$&$0$&$40$&$83$&$2.0$&$0$&$40$&$61$\\
$2.5$&$0$&$50$&$34$&$2.5$&$0$&$50$&$22$\\
$3.0$&$0$&$100$&$2$&$3.0$&$0$&$100$&$1$\\
$3.5$&$0$&$200$&$1$&$3.5$&$0$&$200$&$0$\\
$4.0$&$0$&$300$&$0$&$4.0$&$0$&$300$&$0$\\
$4.5$&$0$&$400$&$0$&$4.5$&$0$&$400$&$0$\\
$5.0$&$0$&$500$&$0$&$5.0$&$0$&$500$&$0$\\
\tableline
\end{tabular}
\end{center}
\end{table*}
In our results, the value of $\left<\mu\right>_{\hat{\cal E}}$ 
is smaller than the magnification in the global universe $\mu_F$. 
Here it should be noted that 
$\left<\mu\right>_{\hat{\cal E}}$ is not an observable quantity in the 
gravitational wave observations, since we can observe 
the only amplitude of superposed waves, although 
it is observable in optical observations, e.g., 
the observation of Type Ia supernovae. 
As mentioned in \S \ref{sec:dispro}, several 
works have discussed 
 the equality 
between $\left<\mu\right>$ and $\mu_F$
~\citep{Weinberg:1976,Ellis:1998ha,Claudel:2000ug,Rose:2001qi,Kibble:2004tm}.
However, in practical sense, $\left<\mu\right>$ 
is not an observable quantity and thus $\left<\mu\right>_{\cal E}$ 
might be more important. 
We will discuss this issue in connection with the optical observation 
elsewhere.

\subsection{Numerical results}\label{sec:nr}
\begin{figure}[htbp]
 \begin{center}
  \includegraphics[scale=0.8]{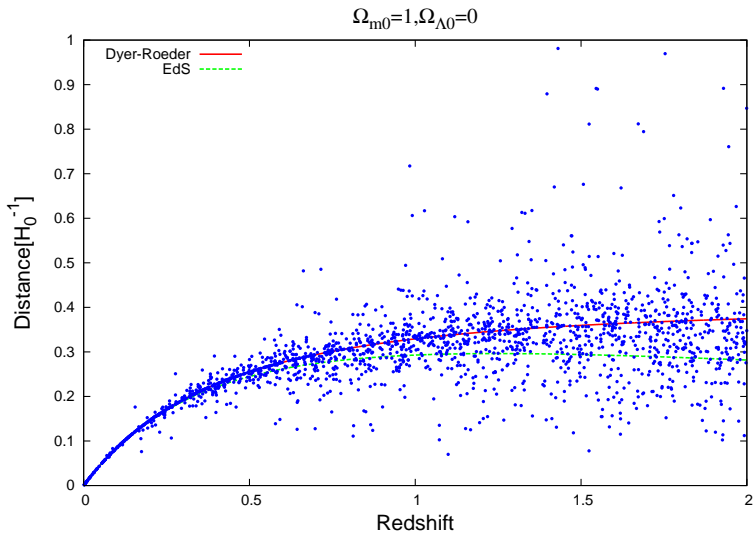}
  \caption{Distance-redshift relation in the case of short
  wavelength $\lambda \ll M_L$. 
  The global geometry is given by the EdS universe. 
  We have numerically generated $2,000$ samples but we cannot plot all of the 
  data points in this figure, since some samples have distances too large to 
  plot within the frame.
  The DR distance is depicted by the solid line and
  the angular diameter distance in the global 
  universe is represented by the dashed line.
  }
  \label{fig:geoeds}
 \end{center}
\end{figure}

\begin{figure}[htbp]
 \begin{center}
  \includegraphics[scale=0.8]{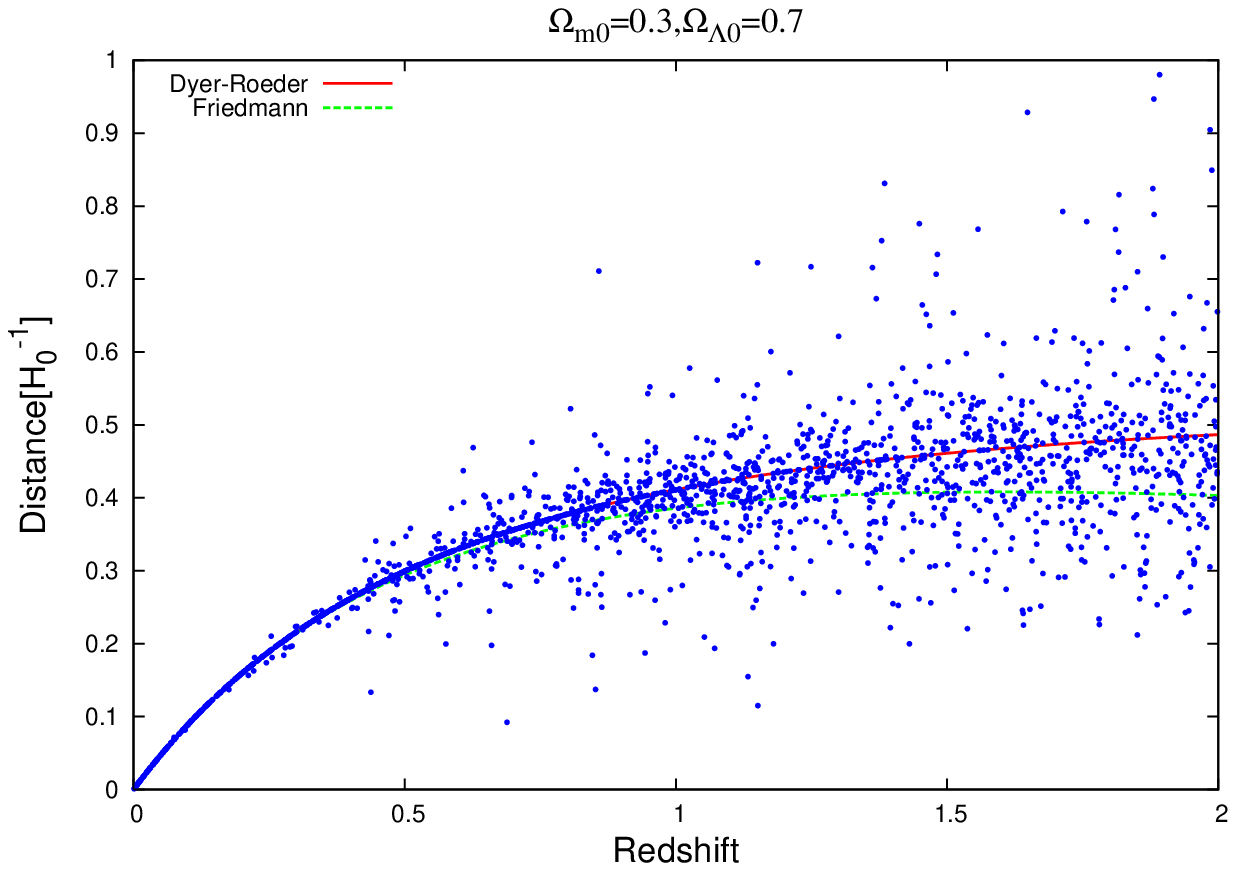}
  \caption{Same as in Fig.~\ref{fig:geoeds}, but here 
  the global geometry is given by the FL universe with 
  $\Omega_{\Lambda0}=0.7,\Omega_{\rm m0}=0.3$. 
  }
\label{fig:geolcdm}
 \end{center}
\end{figure}

\begin{figure}[htbp]
 \begin{center}
  \includegraphics[scale=0.8]{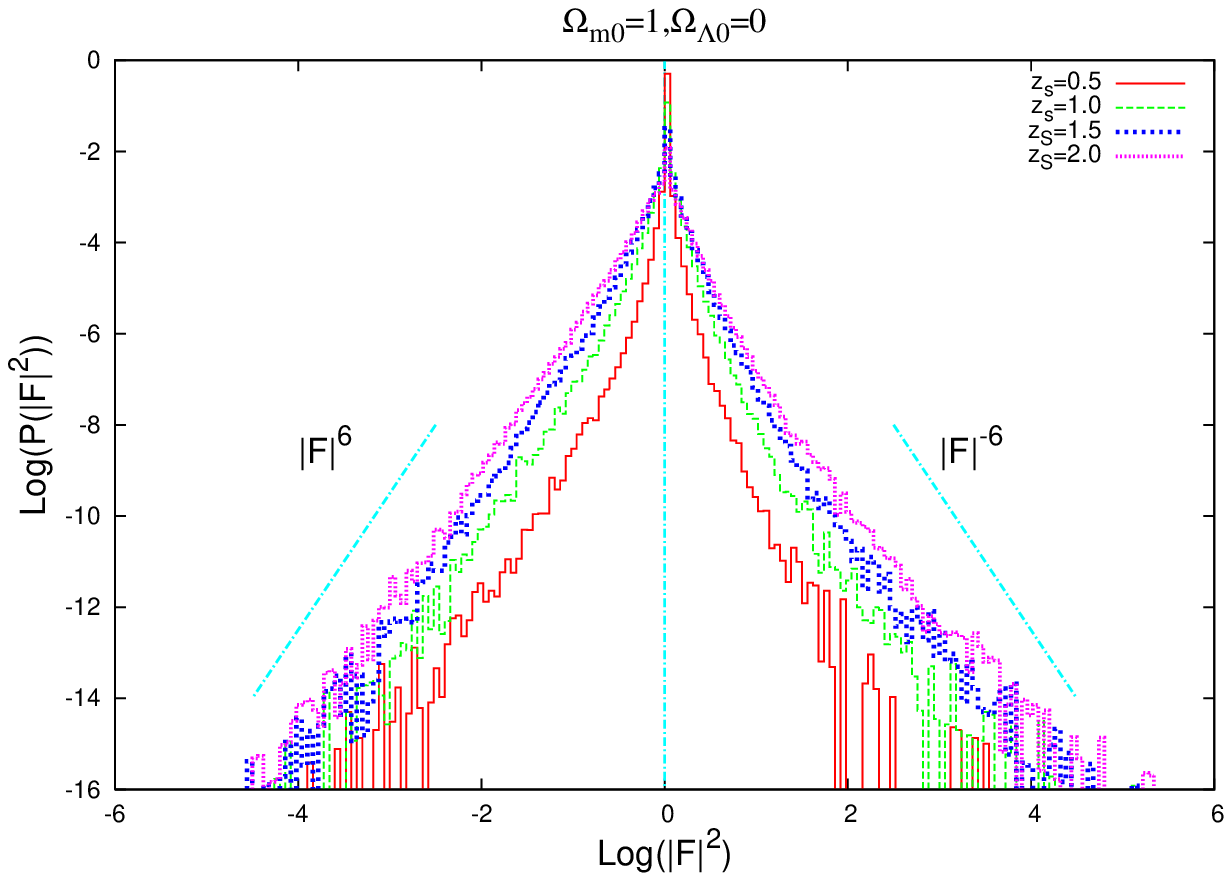}
  \caption{Probability distribution functions $P_{\rm a}(\left|F\right|^2)$ 
in several source redshifts. 
  The global geometry is given by the EdS universe. 
  We have numerically generated $100,000$ samples for each $z_S$.
  }
  \label{fig:histf2eds}
 \end{center}
\end{figure}

\begin{figure}[htbp]
 \begin{center}
  \includegraphics[scale=0.8]{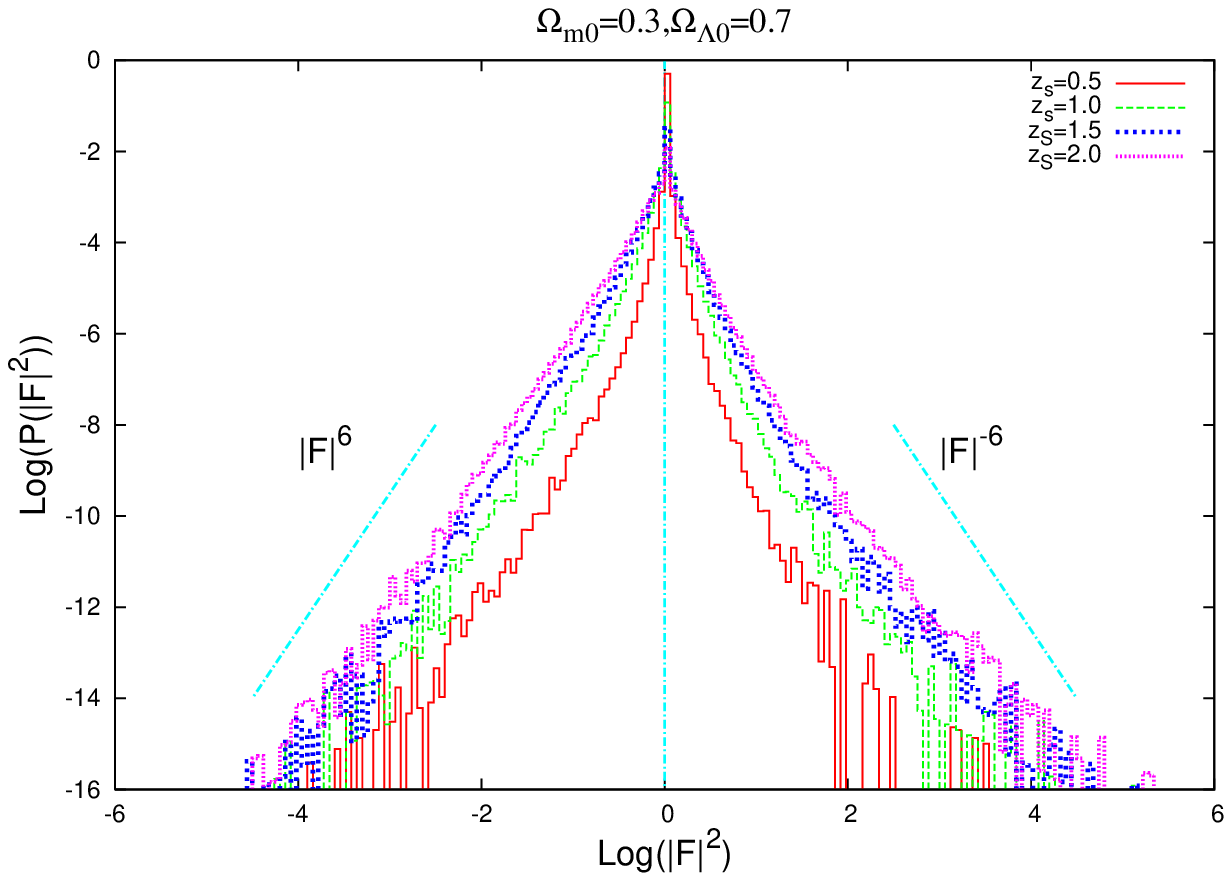}
  \caption{Same as in Fig.~\ref{fig:geoeds}, but here 
  the global geometry is given by the FL universe with 
  $\Omega_{\Lambda0}=0.7,\Omega_{\rm m0}=0.3$. 
  }
\label{fig:histf2lcdm}
 \end{center}
\end{figure}

\begin{figure}[htbp]
\begin{center}
\includegraphics[scale=0.8]{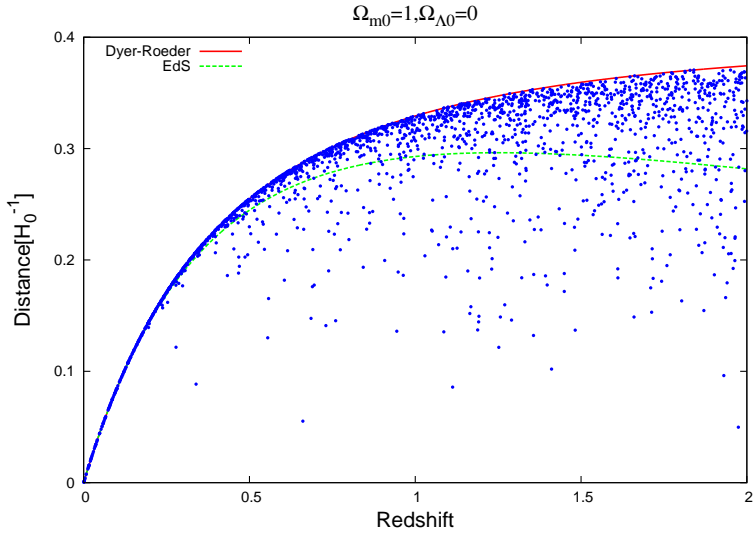}
\caption{Distance-redshift relation in the case of incoherent waves 
with short wavelength $\lambda\ll M_L$. 
We have numerically generated 2000 samples, taking into account 
only the focusing 
effect generated by the first term on the right-hand side of 
eq.~(\ref{eq:ampmag}). 
The DR distance is depicted by the solid line and the angular diameter distance in the 
global universe is represented by the dashed line. 
}
\label{fig:mageds}
\end{center}
\end{figure}

\begin{figure}[htbp]
\begin{center}
\includegraphics[scale=0.8]{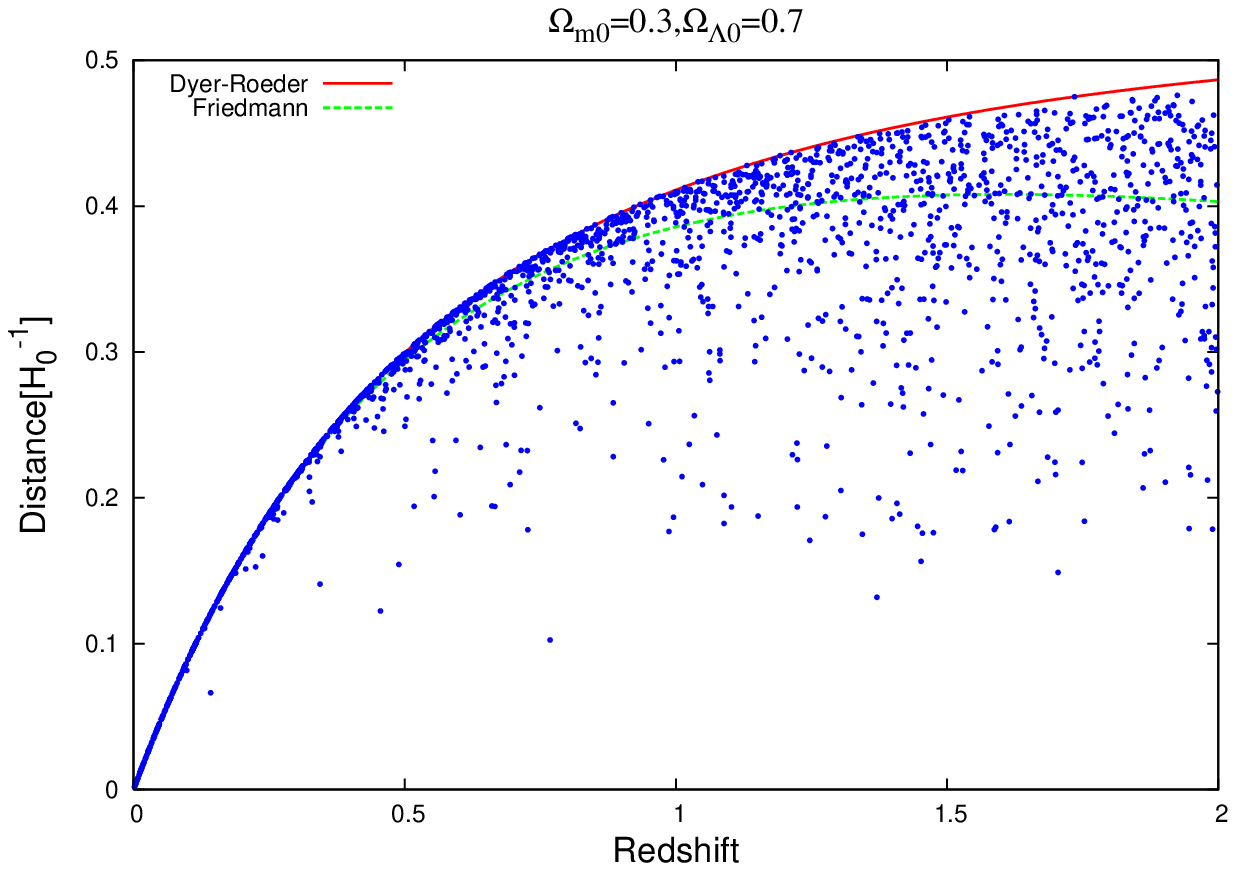}
\caption{Same as in Fig.~\ref{fig:mageds}, but here 
  the global geometry is given by the FL universe with 
  $\Omega_{\Lambda0}=0.7,\Omega_{\rm m0}=0.3$. 
}
\label{fig:maglcdm}
\end{center}
\end{figure}

\begin{figure}[htbp]
 \begin{center}
  \includegraphics[scale=0.8]{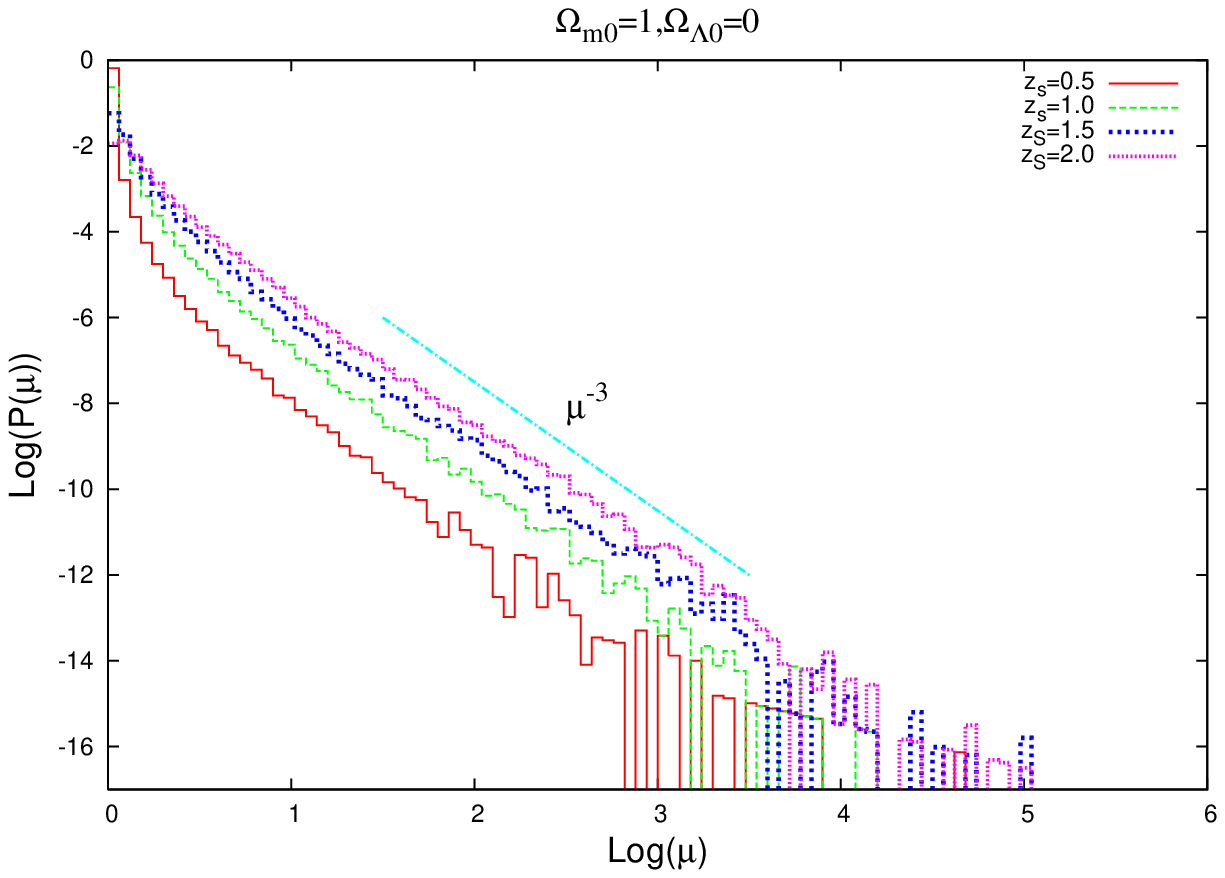}
  \caption{Probability distribution functions $P_{\rm m}(\mu)$ 
in several source redshifts. 
  The global geometry is given by the EdS universe. 
  We have numerically generated $100,000$ samples for each $z_S$.
  }
  \label{fig:histmueds}
 \end{center}
\end{figure}

\begin{figure}[htbp]
 \begin{center}
  \includegraphics[scale=0.8]{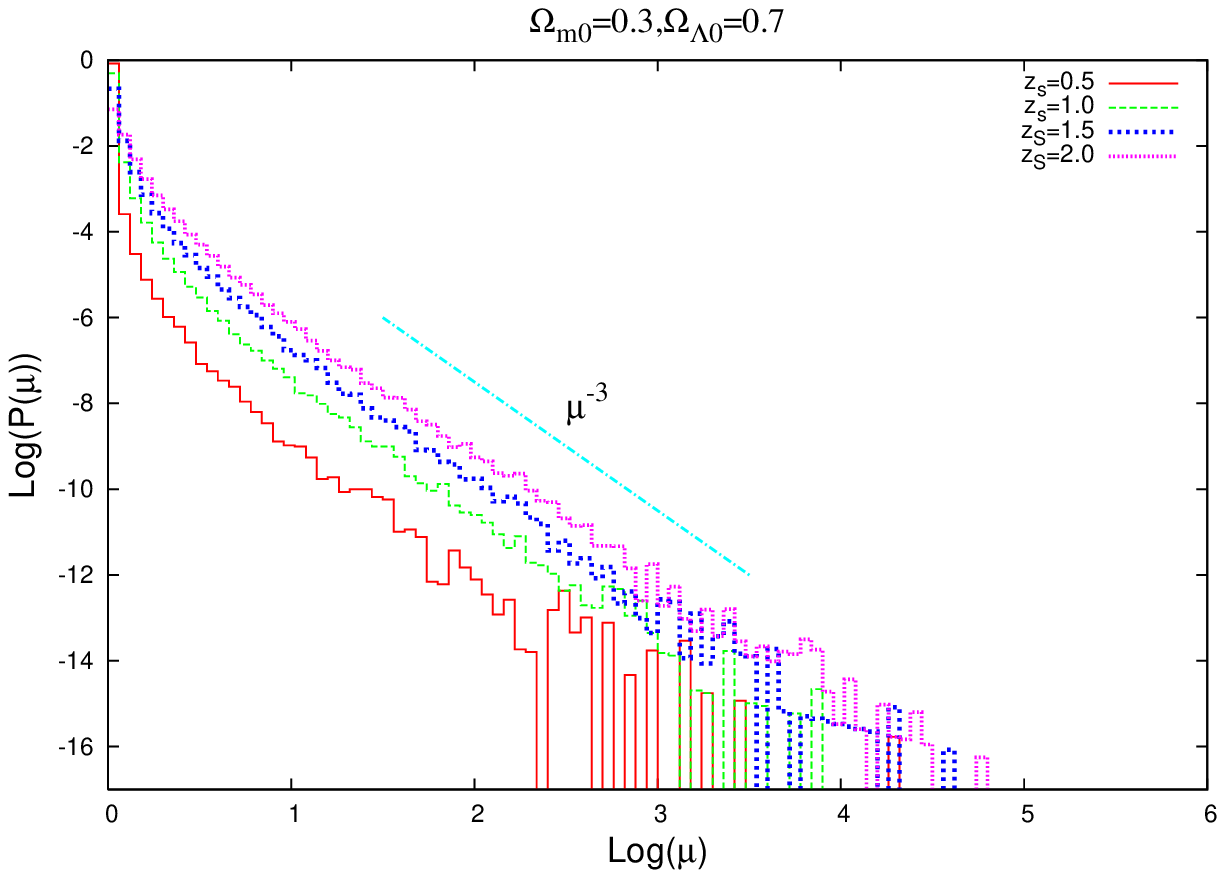}
  \caption{Same as in Fig.~\ref{fig:geoeds}, but here 
  the global geometry is given by the FL universe with 
  $\Omega_{\Lambda0}=0.7,\Omega_{\rm m0}=0.3$. 
  }
\label{fig:histmulcdm}
 \end{center}
\end{figure}

The distance-redshift relation of $2,000$ samples is plotted in 
Figures \ref{fig:geoeds} and \ref{fig:geolcdm}. 
The probability distribution functions $P_{\rm a}(\left|F\right|^2)$ 
in several source 
redshifts are depicted in Figures 
\ref{fig:histf2eds} and \ref{fig:histf2lcdm}. 
The results do not depend on the value of $\lambda/M_L$ if it is much
smaller than unity. 
In the case of $P_{\rm a}(|F|^2)$, the gradient in the high-amplification tail 
is more gradual than $|F|^{-6}$, 
whereas it is approximately proportional to 
$|F|^{6}$ in the low-amplification tail 
(see Figs.~\ref{fig:histf2eds} and \ref{fig:histf2lcdm}). 
For comparison, we show the distance-redshift relation 
for incoherent waves
~(see Figs.~\ref{fig:mageds} and \ref{fig:maglcdm}) and 
the probability distribution function of the magnification $P_{\rm m}(\mu)$
~(see Figs.~\ref{fig:histmueds} and \ref{fig:histmulcdm}). 
The probability of receiving an extremely high magnification 
is very small for each ray. 
Therefore, if a ray is highly magnified,  
it comes from only one lens and the lens is located 
near the line of sight. 
In this case, we find 
\begin{equation}
\mu\propto y^{-1} ~~~{\rm for}~~~y\rightarrow 0
\end{equation} 
from equation (\ref{eq:totmag}). 
Hence, using equation (\ref{eq:p3}), we find
\begin{equation}
P_{\rm m}(\mu)\propto p\left(y(\mu),z\right)\frac{dy}{d\mu}\propto \mu^{-3}. 
\end{equation}
We can see the consistent results in 
Figures \ref{fig:histmueds} and \ref{fig:histmulcdm}. 


The distance dispersion is much larger than for long 
wavelengths, $\lambda\gg M_L$.
The magnification of the amplitude comes from both the
focusing and the interference effects, whereas the demagnification 
comes only from the interference effect.
In contrast to the long-wavelength case, very large magnifications 
due to the focusing effect are possible for 
$\lambda\ll M_L$, because there is no significant diffraction. 
As already mentioned, the first term on the right-hand side of
equation (\ref{eq:ampmag}) includes the shear and 
focusing effects on the ray bundle, 
which can also be found in 
the case of incoherent waves~\citep{Holz:1997ic}, 
and these effects necessarily 
cause magnification of the amplitude~(see Figs.~\ref{fig:mageds} and \ref{fig:maglcdm}). 
By contrast, since the second term on the 
right-hand side of equation (\ref{eq:ampmag}) 
can take either sign, the amplitude can be strongly demagnified, 
i.e., $\left|F_{\rm total}\right|<1$. Thus very large demagnifications 
are also possible for short wavelengths, $\lambda\ll M_L$.  
As the result, the distance determined by the gravitational waves 
can grow longer than the DR distance.
This is a peculiar nature of coherent waves.

\begin{figure}[htbp]
\begin{center}
\includegraphics[scale=0.8]{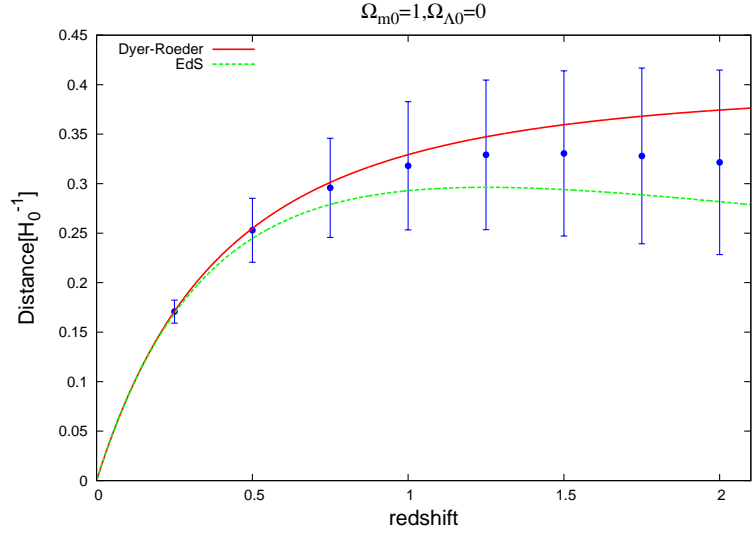}
\caption{Average distance and distance dispersion
 at each redshift in the clumpy universe. 
 The global universe is given by the EdS universe. 
}
\label{fig:deveds}
\end{center}
\end{figure}

\begin{figure}[htbp]
\begin{center}
\includegraphics[scale=0.8]{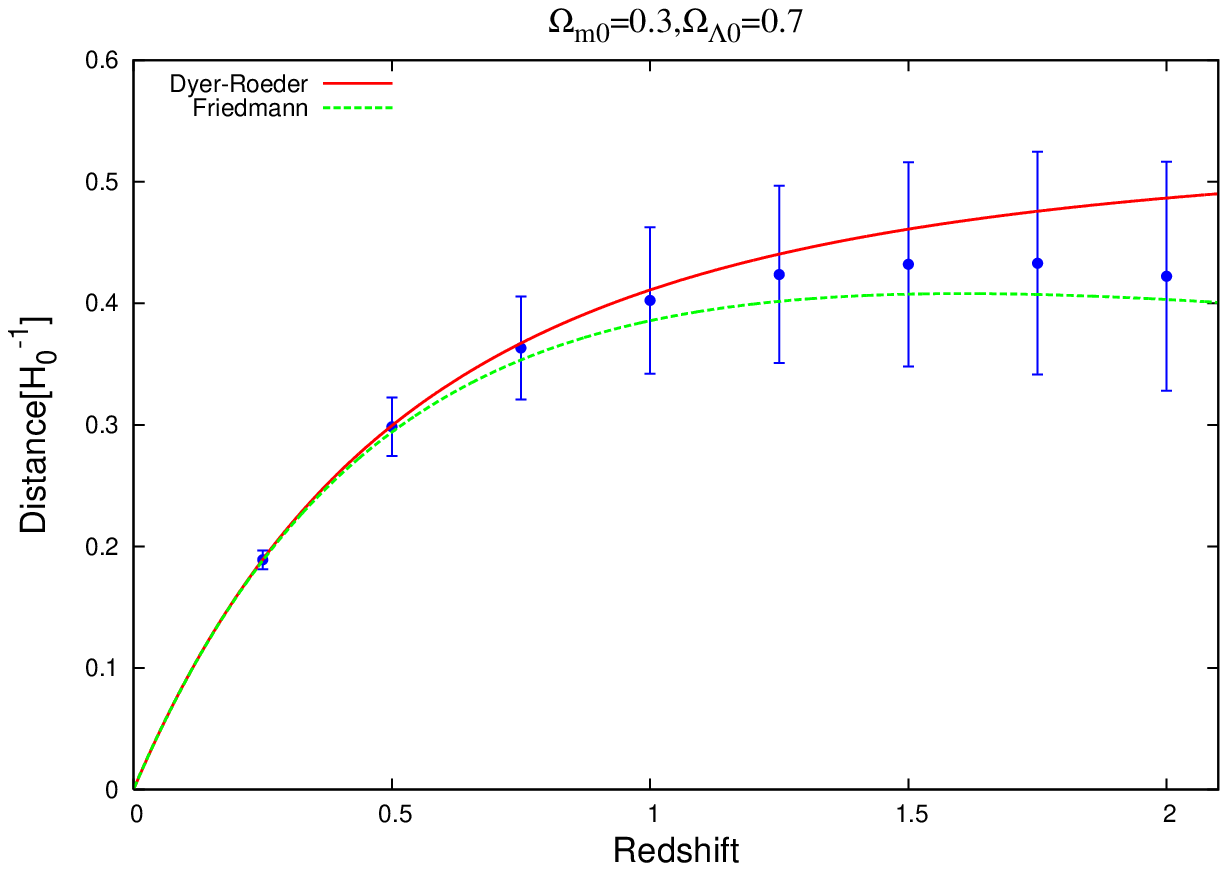}
\caption{Same as in Fig.~\ref{fig:deveds}, but here 
 the global geometry is given by the FL universe with 
 $\Omega_{\Lambda0}=0.7,\Omega_{\rm m0}=0.3$. 
}
\label{fig:devlcdm}
\end{center}
\end{figure}

Here we evaluate the distance dispersion that could be obtained by the data 
provided by gravitational waves if 
our universe is well described by the present 
clumpy mass distribution model. 
For this purpose, we have to take a detection limit for weak 
events into account. 
Let us consider the observation of $1 \mathrm{Hz}$ gravitational waves
emitted by neutron star-neutron star binaries that have the 
characteristic amplitude $h_c$, given by 
\begin{equation}
 h_c
 = 9.88 \times 10^{-24} (1 + z_S)^{5/6}
   \left(%
     \frac{d_{\rm l}}{10 \mathrm{Gpc}}
   \right)^{-1}
   \times\left\{
   \begin{array}{l}
    \frac{1}{2}\left(\cos^2\theta+1\right)\\
    \cos\theta
   \end{array}\right. ,
\end{equation}
where $\theta$ and $d_{\rm l}$ are the inclination angle and 
luminosity distance of the binary~\citep{300}, respectively, 
and we have set the chirp
mass of the binary to $1.2M_\odot$.
The luminosity distance determined by lensed gravitational waves is given
by $(1+z_S)^2D_{DR}(0,z_S)/\left|F_{\rm total}\right|$. 
Assuming the planed sensitivity of 
{\em DECIGO}~\citep{Seto:2001qf}, we set the detection limit as 
\begin{equation}
 h_c > 10^{-23}. 
 \label{eq:limit}
\end{equation}
Randomly choosing the inclination angle $\theta$, 
we numerically generate 
$20,000$ samples of $h_c$ for each source redshift and then select those 
that satisfy equation (\ref{eq:limit}).
We denote the number of detectable samples as $N_{\rm det}$.  
Results are shown in Tables~\ref{tab:deveds} and \ref{tab:devlcdm}
 and Figures \ref{fig:deveds} and \ref{fig:devlcdm}. 
The distance dispersion $\delta D_{\rm obs}$ is much larger than the estimate 
obtained by assuming a weak gravitational field for density perturbation
~\citep{Takahashi:2005ug}. 
This is because \citet{Takahashi:2005ug} 
assumed a linear approximation for the lensed
 waveform even in the short wavelength limit.
Furthermore, he considered the density perturbation as the inhomogeneity, 
but in
 this paper we consider all the matter in the universe to be 
 the point masses,
 which is a more extreme inhomogeneous model.

\begin{table*}[htbp]
\begin{center}
 \caption{Average distance $\left<D_{\rm obs}\right>$ 
 and distance dispersion $\delta D_{\rm obs}$ for each source
 redshift, for EdS universe. The global geometry is given by the EdS universe.
 We have calculated the amplification factors $\left|F_{\rm total}\right|$ 
 of $20,000$ samples for each redshift. 
 \label{tab:deveds}}
\begin{tabular}{ccccc|cc}
\tableline\tableline
$z_S$&$N_{\rm det}$&$\delta D_{\rm obs}[H_0^{-1}]$
&$\delta D_{\rm obs}/\left<D_{\rm obs}\right>$
&$\left<D_{\rm obs}\right>[H_0^{-1}]$&$D_{DR}[H_0^{-1}]$&$D_{F}[H_0^{-1}]$\\
\tableline
0.25&9339&1.16$\times10^{-2}$&6.81\%&1.71$\times10^{-1}$&1.71$\times10^{-1}$&1.69$\times10^{-1}$\\

0.5&8863&3.23$\times10^{-2}$&12.8\%&2.53$\times10^{-1}$&2.55$\times10^{-1}$&2.45$\times10^{-1}$\\

0.75&8450&5.01$\times10^{-2}$&16.9\%&2.96$\times10^{-1}$&3.01$\times10^{-1}$&2.79$\times10^{-1}$\\

1.0&8066&6.48$\times10^{-2}$&20.3\%&3.18$\times10^{-1}$&3.29$\times10^{-1}$&2.93$\times10^{-1}$\\

1.25&7673&7.56$\times10^{-2}$&22.9\%&3.29$\times10^{-1}$&3.47$\times10^{-1}$&2.96$\times10^{-1}$\\

1.5&7221&8.35$\times10^{-2}$&25.2\%&3.31$\times10^{-1}$&3.60$\times10^{-1}$&2.94$\times10^{-1}$\\

1.75&6780&8.88$\times10^{-2}$&27.1\%&3.28$\times10^{-1}$&3.68$\times10^{-1}$&2.89$\times10^{-1}$\\

2.0&6113&9.32$\times10^{-2}$&29.0\%&3.22$\times10^{-1}$&3.74$\times10^{-1}$&2.82$\times10^{-1}$\\
\tableline
\end{tabular}
\end{center}
\end{table*}
\begin{table*}[htbp]
\begin{center}
\caption{Same as Table~\ref{tab:deveds}, but here 
 the global geometry is given by the FL universe 
 with $\Omega_{\Lambda0}=0.7,\Omega_{\rm m0}=0.3$.
\label{tab:devlcdm}}
\begin{tabular}{ccccc|cc}
\tableline\tableline
$z_S$&$N_{\rm det}$&$\delta D_{\rm obs}[H_0^{-1}]$
&$\delta D_{\rm obs}/\left<D_{\rm obs}\right>$
&$\left<D_{\rm obs}\right>[H_0^{-1}]$&$D_{DR}[H_0^{-1}]$&$D_{F}[H_0^{-1}]$\\
\tableline
0.25&9307&7.84$\times10^{-3}$&4.15\%&1.89$\times10^{-1}$&1.89$\times10^{-1}$&1.88$\times10^{-1}$\\

0.5&8711&2.41$\times10^{-2}$&8.07\%&2.98$\times10^{-1}$&2.99$\times10^{-1}$&2.94$\times10^{-1}$\\

0.75&8079&4.23$\times10^{-2}$&11.7\%&3.63$\times10^{-1}$&3.67$\times10^{-1}$&3.53$\times10^{-1}$\\

1.0&7492&6.03$\times10^{-2}$&15.0\%&4.02$\times10^{-1}$&4.11$\times10^{-1}$&3.86$\times10^{-1}$\\

1.25&6862&7.29$\times10^{-2}$&17.2\%&4.24$\times10^{-1}$&4.41$\times10^{-1}$&4.02$\times10^{-1}$\\

1.5&6135&8.41$\times10^{-2}$&19.5\%&4.32$\times10^{-1}$&4.61$\times10^{-1}$&4.08$\times10^{-1}$\\

1.75&5408&9.16$\times10^{-2}$&21.2\%&4.33$\times10^{-1}$&4.76$\times10^{-1}$&4.07$\times10^{-1}$\\

2.0&4696&9.43$\times10^{-2}$&22.3\%&4.22$\times10^{-1}$&4.87$\times10^{-1}$&4.03$\times10^{-1}$\\
\tableline
\end{tabular}
\end{center}
\end{table*}

The distance dispersion given in our results is larger than 
the typical value of observational uncertainties 
evaluated without lensing effects~\citep{Crowder:2005}.
However, 
our evaluation presented in this section is very basic one. 
In order to apply this evaluation to the future observational data, 
we need more detailed information about other experimental 
uncertainties and degeneracies between the situation 
presented here and others. 
We leave this for future works.

\section{Conclusion and summary}
\label{sec:consum}
We have studied the distance-redshift relation obtainable 
through gravitational waves in a clumpy universe.
We have assumed that all of the matter takes the form of uniformly 
distributed point masses with identical mass, $M_L$, 
and that the global geometry is described by the FL universe model. 
The distance from the source to the observer 
in the clumpy universe differs from that in the global universe, 
even if the redshift of the source is identical, 
since gravitational waves are magnified or demagnified by 
the gravitational lensing effect. 
We have numerically calculated the magnification due 
to multiple lensing effects in two extreme cases. 
In the first case, the wavelength $\lambda$ of the gravitational waves
is much longer than the Schwarzschild radius, $2M_L$, of the point-mass 
lens. In the second case, the wavelength $\lambda$ is 
much shorter than $2M_L$.

In the case of long wavelengths, $\lambda \gg M_L$,
the magnification due to a single lens is always very small, but it 
is not negligible. The total magnification in the multiple 
lens system is determined by the accumulation of all of the magnification 
effects due to each lens system, and is not small. 
We found that the relation between the average distance and redshift 
depends on the ratio $\lambda / M_L$, and it approaches the 
distance-redshift relation of the global universe 
as $\lambda / M_L$ increases.
Furthermore, we found that the distance dispersion also depends on the ratio
$\lambda / M_L$, and it approaches zero as $\lambda / M_L$ increases.
The distance-redshift relation in the global universe 
is found even in the clumpy universe if the wavelength of gravitational 
waves is much longer than the Schwarzschild radius of the lens objects.

In the case of short wavelengths, $\lambda \ll M_L$,
the geometrical optics approximation is valid.
In order to calculate the total magnification,
we employed the multiple lens plane method. 
We found that the distance dispersion is larger than that in the 
case of long wavelength. 
The total magnification comes from the focusing and interference 
effects on the ray bundles which emanate from the source to the observer. 
In this case, diffraction effects are negligible and 
thus the large magnifications due to the focusing effects 
on the ray bundles are possible. Moreover, interference effects 
can either enhance 
the magnifications or cause demagnifications.

These lensing effects cause a distance dispersion in the distance-redshift 
relations obtained from gravitational wave data, and we have shown 
its properties. 
In particular, we have clarified the dependence 
of the distance dispersion on the 
wavelength and the mass of the lens object.
Making use of the properties of this distance dispersion,
we can probe the inhomogeneities of our universe. 
The distance dispersion is sensitive to the ratio $M_L / \lambda$, and 
therefore our result suggest that we might use it to gain 
information about the typical masses of clumps  
that cause gravitational lensing effects. 
In order to obtain detailed information about 
the typical masses of lens objects, 
we need a method to calculate 
the case of $M_L\sim \lambda$~\citep{Yamamoto:2003cd}. 

Suppose that we observe gravitational waves from binaries of compact objects 
in a narrow wave band centered at $\lambda$. 
If the large distance dispersion does not appear in the 
distance-redshift relation obtained by these gravitational waves, 
we can impose a restriction on the typical mass scale 
of the clumping matters of $M_L \ll \lambda$.
On the other hand, if a large distance dispersion appears, 
we can say that the mass of clumps $M_L$ is much larger than the 
wavelength of the gravitational waves.

In order to obtain a statistically reliable dispersion,
we must detect a large number of gravitational waves.
{\em DECIGO}'s planned sensitivity is
$2 \times 10^{-24} \rm{Hz}^{-1/2}$ around $1 \rm{Hz}$; 
this means that $10^5$ chirp signals of coalescing binary neutron
stars should be detected in 1 year~\citep{Seto:2001qf}.
Therefore, it should be possible to obtain reliable distance dispersion 
measurements. 
For example, if we observe gravitational waves 
from binary neutron stars in the 
waveband centered at 
$1 \rm{Hz}$, we can determine whether the typical mass scale $M_L$ of the 
dark matter that takes the form of macroscopic compact objects is 
much smaller or much larger than $10^5 M_{\odot}$. 

\acknowledgements
 We are grateful to H.~Ishihara and colleagues in the 
astrophysics and gravity group of Osaka City University for helpful
discussion and criticism.

\end{document}